\pdfoutput=1
\documentclass[11pt,twoside,a4paper,cmspaper,final,collab]{cms-tdr}
\def\svnVersion{b03670e}\def\svnDate{2026/06/15}\def\cmsCernNoTag{CERN-EP-2025-288}\def\cmsCernDate{\today}\def\cmsMessage{Published in Physical Review D as \href{http://dx.doi.org/10.1103/2m87-3dlr}{\doi{10.1103/2m87-3dlr}.}}
\begin{document}\cmsNoteHeader{SUS-24-007}

\ifthenelse{\boolean{cms@external}}{\providecommand{\cmsTable}[1]{#1}}{\providecommand{\cmsTable}[1]{\resizebox{\textwidth}{!}{#1}}}
\ifthenelse{\boolean{cms@external}}{\providecommand{\cmsLeft}{upper\xspace}}{\providecommand{\cmsLeft}{left\xspace}}
\ifthenelse{\boolean{cms@external}}{\providecommand{\cmsRight}{lower\xspace}}{\providecommand{\cmsRight}{right\xspace}}

\newcommand{\baryonicZprime}{baryonic-\PZpr}
\newcommand{\mZpr}{\ensuremath{m_{\PZpr}}\xspace}
\newcommand{\sintheta}{\ensuremath{\sin\theta}\xspace}
\newcommand{\tanbeta}{\ensuremath{\tan\beta}\xspace}
\newcommand{\cossquaretheta}{\ensuremath{\cos^{2}\theta}\xspace}
\newcommand{\lambdathree}{\ensuremath{\lambda_{3}}\xspace}
\newcommand{\lambdaPone}{\ensuremath{\lambda_{P_{1}}}\xspace}
\newcommand{\lambdaPtwo}{\ensuremath{\lambda_{P_{2}}}\xspace}
\newcommand{\twohdma}{2HDM+\Pa}
\newcommand{\ptvecl}{\ensuremath{\vec{p}_{\mathrm{T}}^{\,\smash[b]{\Pell}}}\xspace}
\newcommand{\ptvecjet}{\ensuremath{\vec{p}_{\mathrm{T}}^{\,\text{jet}}}\xspace}
\newcommand{\ptvecakeight}{\ensuremath{\vec{p}_{\mathrm{T}}^{\,\smash[b]{\text{AK8}}}}\xspace}
\newcommand{\ptvecbbbar}{\ensuremath{\vec{p}_{\mathrm{T}}^{\,\smash[b]{\bbbar}}}\xspace}
\newcommand{\recoil}{\ensuremath{U}\xspace}
\newcommand{\vecrecoil}{\ensuremath{\vec{U}}\xspace}
\newcommand{\ma}{\ensuremath{m_{\Pa}}\xspace}
\newcommand{\mA}{\ensuremath{m_{\PSA}}\xspace}
\newcommand{\mh}{\ensuremath{m_{\PSh}}\xspace}
\newcommand{\mH}{\ensuremath{m_{\PH}}\xspace}
\newcommand{\mHpm}{\ensuremath{m_{\PSHpm}}\xspace}
\newcommand{\mchi}{\ensuremath{m_{\PGc}}\xspace}
\newcommand{\vjet}{\ensuremath{\PV\text{+jets}}\xspace}
\newcommand{\Znunujets}{\ensuremath{\PZ(\PGn\PGn)\text{+jets}}\xspace}
\newcommand{\Zlljets}{\ensuremath{\PZ(\Pell\Pell)\text{+jets}}\xspace}
\newcommand{\Wjets}{\ensuremath{\PW\!\text{+jets}}\xspace}
\newcommand{\PAGc}{{\HepAntiParticle{\PGc}{}{}}\xspace}
\newcommand{\PQj}{{\HepParticle{j}{}{}}\xspace}
\newcommand{\chichibar}{\ensuremath{\PGc\PAGc}\xspace}
\newcommand{\pp}{\ensuremath{\Pp\Pp}\xspace}
\newcommand{\gchi}{\ensuremath{g_{\PGc}}\xspace}
\newcommand{\gqua}{\ensuremath{g_{\PQq}}\xspace}
\newcommand{\pth}{\ensuremath{\pt^{\smash[b]{\PSh}}}\xspace}
\newcommand{\mSD}{\ensuremath{m_{\text{SD}}}\xspace}
\newcommand{\Nell}{\ensuremath{N_{\Pell}}\xspace}
\newcommand{\Njets}{\ensuremath{N_{\text{jets}}}\xspace}
\newcommand{\Nbjets}{\ensuremath{N_{\text{bjets}}}\xspace}
\newcommand{\Dphimissjet}{\ensuremath{\Delta\phi(\ptvecmiss,\ptvecjet)}\xspace}
\newcommand{\Dphimissakeight}{\ensuremath{\Delta\phi(\ptvecmiss,\ptvecakeight)}\xspace}
\newcommand{\Dphipftrack}{\ensuremath{\Delta\phi(\text{PF }\ptvecmiss,\text{track }\ptvecmiss)}\xspace}
\newcommand{\topl}{\ensuremath{\PQt(\Pell)}\xspace}
\newcommand{\zll}{\ensuremath{\PZ(\Pell\Pell)}\xspace}
\newcommand{\mbb}{\ensuremath{m_{\bbbar}}\xspace}

\cmsNoteHeader{SUS-24-007}

\title{Search for dark matter produced in association with a Higgs boson decaying to bottom quarks in proton-proton collisions at \texorpdfstring{$\sqrt{s} = 13\TeV$}{sqrt(s) = 13 TeV}}

\date{\today}

\abstract{A search for dark matter particles produced in association with a Higgs boson decaying to a bottom quark-antiquark pair in proton-proton collisions at $\sqrt{s}=13\TeV$ is presented. The data, collected with the CMS detector at the LHC, correspond to an integrated luminosity of 101\fbinv. The analysis is performed in exclusive categories targeting both Lorentz-boosted (merged) and resolved \PQb jet pair topologies, covering a wide range of Higgs boson transverse momentum. A statistical combination is made with a previous search using data collected in 2016 and corresponding to an integrated luminosity of 35.9\fbinv. The observed data agree with the standard model background predictions. Constraints are placed on models predicting new particles or interactions, such as those in the simplified frameworks of \baryonicZprime and \twohdma, where the latter is a type-II two-Higgs-doublet model featuring a heavy pseudoscalar with an additional light pseudoscalar. Upper limits at 95\% confidence level are set on the production cross section for these models. For the \baryonicZprime model, \PZpr boson masses below 2.25\TeV are excluded for a dark matter particle candidate mass of 1\GeV. In the \twohdma model, heavy pseudoscalar masses between 850 and 1300\GeV are excluded for a light pseudoscalar mass of 350\GeV.}

\hypersetup{
pdfauthor={CMS Collaboration},
pdftitle={Search for dark matter produced in association with a Higgs boson decaying to bottom quarks in proton-proton collisions at sqrt(s) = 13 TeV},
pdfsubject={CMS},
pdfkeywords={CMS, dark matter, mono-Higgs, bottom quarks}}

\maketitle

\section{Introduction}

Strong astrophysical evidence supports the existence of dark matter (DM), providing compelling motivation for physics beyond the standard model (SM). Cosmological observations indicate that DM constitutes approximately 85\% of the matter content in the universe~\cite{Planck:2018nkj}. Since the existing evidence is based only on its gravitational interaction, whether DM interacts with the SM particles remains an open question. Many beyond the SM theories suggest a particle nature of DM~\cite{ref:DMNPBSM}, proposing several particle candidates that are compatible with the observed relic density of DM in the universe~\cite{ref:relic}. The hypothesis that DM consists of stable or long-lived weakly interacting massive particles~\cite{ref:WIMPs,Jungman:1995df} drives many of the experimental searches at the CERN LHC.

A DM particle, if produced in high-energy proton-proton (\pp) collisions at the LHC, can exit the detector undetected, becoming a source of large missing transverse momentum (\ptmiss). The presence of such a weakly interacting particle can be inferred if it is produced in association with a visible SM particle \PX (quark, gluon, photon, or heavy boson). Such an interaction may result in a ``mono-\PX'' signature with an SM particle or jet recoiling against \ptmiss. The SM particle \PX could be emitted directly from a parton as initial-state radiation or from a DM particle via a new coupling to SM particles.

The discovery of the SM Higgs (\PSh) boson~\cite{ATLAS-Higgs,CMS-Higgs1,CMS-Higgs2} allows the additional possibility of probing DM in the mono-Higgs channel~\cite{Carpenter:2013xra, Berlin:2014cfa}, complementing other mono-\PX searches. The production of a Higgs boson from initial-state radiation is highly suppressed because of the mass dependence of its coupling strength to fermions and its loop-suppressed coupling to gluons~\cite{Carpenter:2013xra, Berlin:2014cfa, Petrov:2013nia}. The associated production of DM with a Higgs boson could therefore directly probe the structure of the effective DM-SM coupling~\cite{Carpenter:2013xra}. In addition, there exist models where the coupling between the new physics sector and the Higgs boson is enhanced~\cite{BERTONE2005279,BURGESS2001709} or where the DM particles interact with the SM particles exclusively through the Higgs sector~\cite{John_March_Russell_2008}. Owing to these features, the mono-Higgs channel offers a promising signature for the search for DM.

\begin{figure}[!ht]
  \centering
  \includegraphics[width=0.42\textwidth]{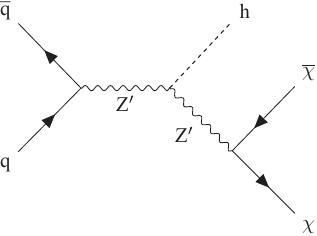}
  \hfill
  \includegraphics[width=0.48\textwidth]{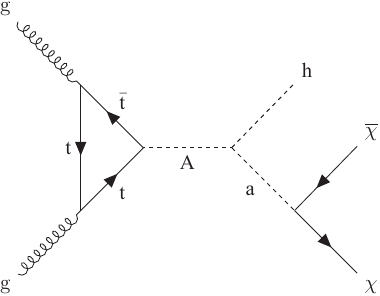}
  \caption{Feynman diagrams for simplified benchmark models considered in this analysis: the \baryonicZprime model (\cmsLeft) and \twohdma model (\cmsRight)~\cite{Abercrombie:2015wmb,abe2018lhc}.}
  \label{fig:feynman_2HDMa_ZpB}
\end{figure}

This analysis considers two simplified benchmark models for DM+Higgs boson production~\cite{Abercrombie:2015wmb,abe2018lhc}. Figure~\ref{fig:feynman_2HDMa_ZpB} shows representative Feynman diagrams corresponding to these models.

In the \baryonicZprime model~\cite{Carpenter:2013xra}, the \PZpr particle is a vector boson mediator corresponding to a new baryon number symmetry, $U(1)_{\text{B}}$. It radiates an \PSh boson before decaying to a pair of Dirac fermions, \chichibar, which serve as DM candidates. The strength of the coupling between the \PSh and \PZpr bosons is assumed to be directly proportional to the mediator mass \mZpr. The coupling strengths of the \PZpr boson are chosen following Refs.~\cite{Abercrombie:2015wmb,abe2018lhc}, with the coupling between the \PZpr boson and DM particles, $\gchi=1$, and between the \PZpr boson and quarks, $\gqua=0.25$. The shape of the \ptmiss distribution depends mainly on two parameters, \mZpr and the mass of the DM candidate \mchi, which are free parameters in the model and are varied in this search. This paper considers \mZpr between 100 and 3500\GeV, and \mchi between 1 and 800\GeV.

The \twohdma model~\cite{Bauer2017} is an extension of the type-II two-Higgs-doublet model~\cite{BRANCO20121}, introducing an additional light pseudoscalar boson, \Pa. The model predicts six mass eigenstates: the light scalar Higgs boson (\PSh, mass \mh), the heavy scalar Higgs boson (\PH, mass \mH), the light pseudoscalar (\Pa, mass \ma), the heavy pseudoscalar (\PSA, mass \mA), and two heavy charged scalars (\PSHpm, mass \mHpm). The states \PSh and \PH are linear combinations of the $CP$-even weak eigenstates with mixing angle $\alpha$, while the \Pa and \PSA states are linear combinations of the $CP$-odd weak eigenstates with mixing angle $\theta$. The ratio of the vacuum expectation values of the two Higgs doublets is \tanbeta. The \Pa boson decays to a pair of Dirac fermions, \chichibar.  The model also includes three quartic couplings: \lambdathree, \lambdaPone, and \lambdaPtwo. Constraints from perturbativity and unitarity restrict their magnitude and sign, and benchmark values are chosen as $\lambdathree=\lambdaPone=\lambdaPtwo=3$~\cite{Abercrombie:2015wmb,abe2018lhc,Bauer2017}. The parameter scans and fixed values are described in Section~\ref{subsec:2HDMa_results}. An important advantage of the \twohdma framework over simpler models is its rich phenomenology, which can yield complementary exclusion limits from different experimental searches.

This paper presents a search for DM produced in association with a Higgs boson decaying to a bottom quark-antiquark pair (\bbbar). This \bbbar decay mode has the largest branching fraction and provides the highest signal yield among possible decay channels. A previous search in the same final state by the CMS experiment~\cite{exo:monohbb2016} was based on data collected in 2016 at a center-of-mass energy $\sqrt{s}=13\TeV$ with an integrated luminosity of 35.9\fbinv. A similar search has been published by the ATLAS experiment at $\sqrt{s}=13\TeV$ corresponding to an integrated luminosity of 139\fbinv~\cite{ATLAS-monoHiggsBB}. No significant excess of events over the SM prediction was observed in either study.

The final state for this analysis, mono-$\PSh(\bbbar)$, is characterized by the presence of high \ptmiss and a Higgs boson decaying to a \bbbar pair. The analysis is broadly categorized into merged and resolved categories, based on the transverse momentum of the reconstructed \PSh boson candidate, \pth. Events with large \pth are assigned to the merged category, while those with relatively low to moderate \pth are included in the resolved category.

Compared to Ref.~\cite{exo:monohbb2016}, we have nearly tripled the amount of data analyzed and enhanced the analysis sensitivity by means of improved identification of $\PSh\to\bbbar$ decays. While such decays were previously selected using $N$-subjettiness~\cite{Thaler:2010tr}, we now use criteria based on a deep neural network (DNN), resulting in increased sensitivity to the merged signals. We have further extended the sensitivity by combining the new results with the 2016 results, yielding a total data set of 138\fbinv, equivalent in size to that of Ref.~\cite{ATLAS-monoHiggsBB}.

This paper is organized as follows. Section~\ref{sec:Detector} provides a brief introduction to the CMS detector. Section~\ref{sec:Data} describes the data and simulation samples used. Section~\ref{sec:EventReco} discusses the event reconstruction and various object definitions used in the search. The event selection, background estimation, and signal extraction procedures are detailed in Section~\ref{sec:EventSel}. Systematic uncertainties, both experimental and theoretical, are discussed in Section~\ref{sec:Syst}. The results are presented in Section~\ref{sec:Result}, and a summary is given in Section~\ref{sec:Summary}. Tabulated results are provided in the HEPData record for this analysis~\cite{hepdata}.

\section{The CMS detector}
\label{sec:Detector}

The central feature of the CMS apparatus is a superconducting solenoid with internal diameter of 6\unit{m}, providing a magnetic field of 3.8\unit{T}. Within the solenoid volume are a silicon pixel and strip tracker, a lead tungstate crystal electromagnetic calorimeter (ECAL), and a brass and scintillator hadron calorimeter (HCAL), each composed of a barrel and two endcap sections. Forward calorimeters extend the pseudorapidity ($\eta$) coverage provided by the barrel and endcap detectors. Muons are reconstructed in gas-ionization detectors embedded in the steel flux-return yoke outside the solenoid. More detailed descriptions of the CMS detector, together with a definition of the coordinate system used and the relevant kinematic variables, can be found in Refs.~\cite{CMS:2008xjf,CMS:2023gfb}.

Events of interest are selected using a two-tiered trigger system. The first level (L1), composed of custom hardware processors, uses information from the calorimeters and muon detectors to select events at a rate of around 100\unit{kHz} within a fixed latency of 4\mus~\cite{CMS:2020cmk}. The second level, known as the high-level trigger (HLT), consists of a farm of processors running a version of the full event reconstruction software optimized for fast processing, and reduces the event rate to around 1\unit{kHz} before data storage~\cite{CMS:2016ngn}.

\section{Data and simulated samples}
\label{sec:Data}

The data analyzed in this search are from \pp collisions at $\sqrt{s}=13\TeV$ collected by the CMS experiment in 2017 and 2018, corresponding to an integrated luminosity of 101\fbinv. Signal and background contributions are modeled using Monte Carlo (MC) simulations. All samples use NNPDF3.1 parton distribution functions (PDFs) at next-to-next-to-leading order accuracy~\cite{NNPDF:2017mvq, Khachatryan:2015pea}.  The parton shower, hadronization, and the underlying event simulation are provided by \PYTHIA~\cite{pythia8} version 8.202 or later with the underlying event tune CP5~\cite{CMS:2019csb}. The response of the CMS detector is simulated with the \GEANTfour package~\cite{geant4}. Simulated events are reconstructed with the same software as used for data events. All samples include the effects of additional \pp interactions within the same or nearby bunch crossings (pileup). Minimum bias interactions simulated with \PYTHIA are overlaid on the hard scattering interaction with a multiplicity distribution matching that observed in the collider data.

For both the simplified DM models considered, \baryonicZprime and \twohdma, the signal samples are generated at leading order (LO) using \MGvATNLO~\cite{amcatnlo} version 2.6.5. The dominant background arises from the production of \Znunujets and \ttbar pairs. The \Znunujets process is generated using \MGvATNLO at next-to-LO (NLO) precision in perturbative quantum chromodynamics (pQCD). The lepton+jets decay of the \ttbar pair constitutes the leading background contribution compared to dileptonic or hadronic decays. All three \ttbar decay modes are simulated at NLO in pQCD using the {\POWHEG}v2.0 generator~\cite{Nason:2004rx, Frixione:2007vw, Alioli:2010xd}. The \pt spectra of top quarks in \ttbar data have been observed to be significantly softer compared to MC simulations at LO or NLO, attributed to missing higher-order calculations. To mitigate this discrepancy, \ttbar events in simulation are reweighted using a parametric function~\cite{CMS:2018htd}.

Subleading contributions arise from \Wjets, single top quark, and diboson ($\PW\PW$, $\PW\PZ$, and $\PZ\PZ$) processes. The \Wjets background contributes to this analysis when the lepton is not reconstructed or identified, or falls outside the detector acceptance. Events from \Wjets and \Zlljets processes are simulated using the \MGvATNLO generator at LO, and the cross section is scaled to NLO accuracy in pQCD and electroweak (EW) corrections. Event generation for the production of a single top quark in the $t$ channel~\cite{Frederix:2012dh} and in association with a \PW boson~\cite{Re:2010bp} is performed at NLO with the \POWHEG generator. Diboson background events are simulated at NLO in pQCD with the \POWHEG or \MGvATNLO generator.

The SM backgrounds with genuine Higgs bosons are produced via different processes: gluon-gluon fusion, vector boson fusion, in association with a vector boson, or in association with a pair of top quarks. These SM Higgs boson samples are all generated with \POWHEG at NLO in pQCD. The background from SM events composed uniquely of jets produced through the strong interaction, referred to as QCD multijet events, is also considered. These are generated at LO using \MGvATNLO.

The merging schemes employed for the \MGvATNLO LO and NLO samples with the parton shower model are MLM~\cite{Alwall:2007fs} and FxFx~\cite{Frederix:2012ps}, respectively.

\section{Event reconstruction}
\label{sec:EventReco}

The CMS particle-flow (PF) algorithm~\cite{CMS:2017yfk} aims to reconstruct and identify each individual particle in an event, with an optimized combination of information from the various elements of the CMS detector. The energy of photons is obtained from the ECAL measurement. The energy of electrons is determined from a combination of the electron momentum at the primary interaction vertex as determined by the tracker, the energy of the corresponding ECAL cluster, and the energy sum of all bremsstrahlung photons spatially compatible with originating from the electron track. The energy of muons is obtained from the curvature of the corresponding track. The energy of charged hadrons is determined from a combination of their momentum measured in the tracker and the matching ECAL and HCAL energy deposits, corrected for the response function of the calorimeters to hadronic showers. Finally, the energy of neutral hadrons is obtained from the corresponding corrected ECAL and HCAL energies.

The primary vertex (PV) is taken to be the vertex corresponding to the hardest scattering in the event, evaluated using tracking information alone, as described in Section 9.4.1 of Ref.~\cite{CMS-TDR-15-02}.

The electron momentum is estimated by combining the energy measurement in the ECAL with the momentum measurement in the tracker~\cite{CMS:2020uim,CMS-DP-2020-021}. The momentum resolution for electrons with $\pt\approx45\GeV$ from $\PZ\to\Pe\Pe$ decays ranges from 1.6 to 5\%. A multivariate approach is used to identify prompt electrons arising from the decays of \PW or \PZ bosons. Two identification criteria, corresponding to the expected electron identification efficiency of either 90\% (loose working point, WP) or 80\% (tight WP), are used in this analysis. The loose WP is used for rejecting events for the 0-lepton signal region (SR), as well as for rejecting events containing extra electrons for the single or dimuon control regions (CRs). The tight WP is used to select events in the single or dielectron control regions. The electron \pt threshold required for the loose (tight) selection is 10 (35)\GeV.

Muons are measured in the pseudorapidity range $\abs{\eta}<2.4$, with detection planes made using three technologies: drift tubes, cathode strip chambers, and resistive plate chambers. The single muon trigger efficiency exceeds 90\% over the full $\eta$ range, and the efficiency to reconstruct and identify muons is greater than 96\%. Matching muons to tracks measured in the silicon tracker results in a relative transverse momentum resolution, for muons with \pt up to 100\GeV, of 1\% in the barrel and 3\% in the endcaps. In the central region of the detector, the \pt resolution is better than 7\% for muons with \pt up to 1\TeV~\cite{CMS:2018rym}. Two sets of selection conditions, corresponding to 96--98\% (tight WP) and $\sim$99\% (loose WP) efficiency for the identification of prompt muons, are used in the analysis. The use of loose and tight WPs across different analysis regions is similar to the approach used for electrons. The muon \pt threshold for the loose (tight) selection is 15 (30)\GeV.

The contribution of misidentified electrons or muons is further reduced by requiring the lepton candidate to satisfy isolation criteria that limit the total energy of tracks and calorimeter clusters measured in conical regions around them.

{\tolerance=800
Hadronic \PGt decays (\tauh) are reconstructed from jets using the hadrons-plus-strips algorithm~\cite{CMS:2018jrd}, which combines 1 or 3 tracks with energy deposits in the calorimeters to identify the tau decay modes. To distinguish genuine \tauh decays from jets originating from the hadronization of quarks or gluons, and from electrons or muons, a convolutional discriminator based on a DNN called \textsc{DeepTau}~\cite{CMS:2022prd} is used. Information from all individual reconstructed particles near the \tauh axis is combined with properties of the \tauh candidate and the event. The rate of a jet to be misidentified as \tauh by the \textsc{DeepTau} algorithm depends on the \pt and quark flavor of the jet. In simulated events from \PW boson production in association with jets it has been estimated to be 0.43\% for a genuine \tauh identification efficiency of 70\%.
\par}

Jets are reconstructed from the PF particles using the infrared- and collinear-safe anti-\kt algorithm~\cite{Cacciari:2008gp, Cacciari:2011ma} with distance parameters 0.4 (``AK4 jets'') and 0.8 (``AK8 jets''). The distance parameter value 0.4 is found to optimize the reconstruction of jets initiated by a single quark or gluon. For the boosted topology in this analysis, the two \PSh-daughter \PQb jets have angular separation too small to be resolved, but are captured in a single wider jet reconstructed by setting the algorithm's distance parameter to 0.8. Jet momentum is determined as the vectorial sum of all particle momenta in the jet, and is found from simulation to be, on average, within 5 to 10\% of the momentum at particle level over the entire \pt spectrum and detector acceptance.

The pileup-per-particle identification algorithm~\cite{Sirunyan:2020foa,Bertolini:2014bba} is used to mitigate the effect of pileup at the reconstructed particle level, making use of local shape information, event pileup properties, and tracking information. Charged particles identified to be originating from pileup vertices are discarded. For each neutral particle, a local shape variable is computed using the surrounding charged particles compatible with the PV within the tracker acceptance, and using both charged and neutral particles in the region outside of the tracker coverage. The momenta of the neutral particles are then rescaled according to their probability to originate from the PV deduced from the local shape variable, superseding the need for jet-based pileup corrections~\cite{Sirunyan:2020foa}.

Jet energy corrections are derived from simulation studies so that the average measured energy of jets becomes consistent with that of particle-level jets within uncertainties. In situ measurements of the momentum balance in dijet, photon+jet, {\PZ}+jet, and multijet events are used to determine any residual differences between the jet energy scale (JES) and jet energy resolution (JER) in data and in simulation, and appropriate corrections are made~\cite{CMS:2016lmd,CMS-DP-2021-033}. The JER amounts typically to 15--20\% at 30\GeV, 10\% at 100\GeV, and 5\% at 1\TeV~\cite{CMS:2016lmd}. Additional selection criteria are applied to each jet to remove jets potentially stemming from instrumental effects or reconstruction failures~\cite{Sirunyan:2020foa}. Jets overlapping with the selected electrons and muons (\ie those with $\DR<0.4$ for $\DR=\sqrt{\smash[b]{(\Delta\eta)^2+(\Delta\phi)^2}}$ for AK4 jets, and $\DR<0.8$ for AK8 jets) are removed. Only jets within the tracker volume, with $\abs{\eta}<2.5$, are considered in the analysis.

The missing transverse momentum vector \ptvecmiss is computed as the negative vector sum of the transverse momenta of all the PF candidates in an event, and its magnitude is denoted as \ptmiss~\cite{CMS:2019ctu}. The \ptvecmiss is modified to account for corrections to the JES and JER of the reconstructed jets in the event. Anomalous high-\ptmiss events can arise from a variety of reconstruction failures, detector malfunctions or noncollision backgrounds. Such events are rejected by event filters that are designed to identify more than 85--90\% of the spurious high-\ptmiss events with a false positive rate below 0.1\%~\cite{CMS:2019ctu}.

Large-radius AK8 jets are used to identify the Lorentz-boosted hadronically decaying Higgs bosons. For these jets, a grooming algorithm is used, where the constituents of the AK8 jet are reclustered using the Cambridge-Aachen algorithm~\cite{Dokshitzer:1997in,Wobisch:1998wt}. The ``modified mass drop tagger'' algorithm~\cite{Dasgupta:2013ihk,Butterworth:2008iy}, also known as the ``soft-drop'' algorithm~\cite{Larkoski:2014wba}, with angular exponent $\beta=0$, soft cutoff threshold $z_{\text{cut}}<0.1$, and characteristic radius $R_{0}=0.8$, is applied to remove soft, wide-angle radiation from the jet, mitigating pileup and underlying event contamination to improve the mass resolution and enhance the identification of boosted, hadronically decaying Higgs bosons.

The AK4 \PQb jet identification is performed using the \textsc{DeepJet} algorithm~\cite{Bols_2020, Sirunyan_2018}, a DNN-based tagger that utilizes information from charged and neutral particles, as well as secondary vertices within a jet, to determine the likelihood that the jet originates from a bottom quark. For an AK4 jet to be tagged as a bottom jet, its \textsc{DeepJet} \PQb tagging discriminator score is required to be larger than a specific threshold. The threshold values, corresponding to false positive rates for jets initiated by light quarks or gluons of 10, 1, and 0.1\%, are referred to as loose, medium, and tight WPs, respectively. In the resolved category, to identify the \PQb jets originating from the Higgs boson decay, a medium WP is used, which has a signal efficiency of 80\%. In the merged category, to identify and veto AK4 \PQb jets away from the Higgs boson candidate AK8 jet, a loose WP for the \textsc{DeepJet} tagger with signal efficiency of 90\% is used. To match the efficiency of \PQb tagging in data, correction factors are applied to simulated events as a function of jet \pt, $\eta$, and the flavor of the initiating parton. The performance of the \textsc{DeepJet} tagger and its calibration results using 2016--2018 data are summarized in Ref.~\cite{CMS-DP-2023-005}.

To identify AK8 jets originating from the $\PSh\to\bbbar$ decay, a dynamic graph convolutional neural-network-based tagger, \textsc{ParticleNet}~\cite{PhysRevD.101.056019} with its mass decorrelated version (\textsc{ParticleNet-MD}), is used. The algorithm is trained to identify two-prong hadronic decays of Lorentz-boosted heavy particles, and the tagger performance is decorrelated from the mass of the parent particle, which allows it to be used in various analyses with a wide range of resonance masses. It uses jet constituents and secondary vertices associated with the jet, with the tagger score specifying the likelihood that the jet originates from a heavy particle decaying to a \PQb quark-antiquark pair $\PX\to\bbbar$, as compared to light quarks or gluons. The loose WP threshold is used, such that the discriminator score is chosen to achieve a signal efficiency of 80\% and a misidentification rate of 1\% for light quark or gluon jets. Calibration is performed~\cite{CMS-DP-2022-005} using an event sample in data enriched in jets originating from $\Pg\to\bbbar$ splittings, which have a similar substructure as $\PSh\to\bbbar$ decay. Correction factors are applied in simulation, parametrized as functions of the AK8 jet \pt, to match the efficiency of \PQb tagging in data, with a performance summarized in Ref.~\cite{CMS-PAS-BTV-22-001}.

A section of the HCAL was not operational during a portion of the 2018 data-taking period, which affected approximately 65\% of the total integrated luminosity collected in 2018. This resulted in irrecoverable energy mismeasurements in a localized region of the detector, defined by $-1.57<\phi<-0.87$ and $-3.0<\eta<-1.3$. To mitigate the impact of these mismeasurements, events are vetoed if any AK4 jet with $\pt>30\GeV$, or any AK8 jet with $\pt>200\GeV$, is found within the affected region in the 2018 data set. Additionally, in the electron control regions, events are rejected if any electron with $\pt>30\GeV$ is present in the affected region. Events with severe mismeasurement, where a jet is fully lost in the affected HCAL region, are found to contribute predominantly at low \ptmiss ($<$470\GeV) and exhibit a characteristic $\phi(\ptvecmiss)$ distribution. To veto such events, an additional selection is applied: events with $\ptmiss<470\GeV$ are rejected if $-1.62<\phi(\ptvecmiss)<-0.62$. The threshold of 470\GeV is chosen to efficiently remove the majority of these background events, while maintaining high signal efficiency.

\section{Event selection}
\label{sec:EventSel}

In the boosted regime, the $\PSh\to\bbbar$ decay products are highly collimated and reconstructed as an AK8 jet. For low to moderate \pth with well-separated decay products, AK4 jets are used to reconstruct particles from the \PQb quark fragmentation.

The two categories are orthogonal by construction. Events are first considered for the merged category, and those failing this selection are passed on to the selection criteria for the resolved category. This helps improve the overall sensitivity of the analysis to different regions of parameter space.

Events with any loosely reconstructed tau lepton candidate with $\pt>20\GeV$, $\abs{\eta}<2.3$ and $\abs{d_{z}}<0.2\unit{cm}$ are vetoed in all analysis regions, where $d_{z}$ is the longitudinal impact parameter. Events containing an isolated photon candidate with $\pt>20\GeV$, $\abs{\eta}<2.5$, and passing loose identification criteria with 90\% efficiency, are also vetoed in all analysis regions.

\subsection{Signal region}
\label{subsection:SR}

The events in the signal region are selected using a trigger with a \ptmiss threshold of 120\GeV, where the calculation of \ptmiss is based on all PF candidates reconstructed at the HLT. The \pt of muon candidates identified by the PF algorithm is subtracted in this calculation. As a result, events with high-\pt muons are associated with large \ptmiss, which allows us to use the same trigger path to select events populating single-muon and dimuon control regions. The \ptmiss trigger efficiency is around 90\% for $\ptmiss=200\GeV$ and reaches 99\% for 300\GeV and above. Events in the SR are required to have $\ptmiss>250$ (200)\GeV in the merged (resolved) category.

To reduce the \ttbar and \Wjets background contribution in the SRs, events are required to have no isolated electron (muon) with $\pt>10$ (15)\GeV and $\abs{\eta}<2.5$ (2.4) that passes loose quality criteria. The symbol \Pell denotes either an electron or muon, and \Nell the number of electrons or muons in an event.

In the merged-category SR, exactly one AK8 jet ($N_{\text{AK8}}=1$) is required with $\pt>200\GeV$, $\abs{\eta}<2.5$, and soft-drop-corrected mass ($\mSD$) within $70<\mSD<160\GeV$, a window around the \PSh boson mass that includes sidebands for the evaluation of background. To select events where the AK8 jet originates from an $\PSh\to\bbbar$ decay, the jet must pass the loose WP of the \textsc{ParticleNet-MD} double-\PQb algorithm. Furthermore, to reduce the \ttbar background contribution, a selection is placed on the number of AK4 jets, \Njets, where we allow up to two such jets with $\pt>30\GeV$ that do not overlap with the \PSh candidate AK8 jet, \ie, $\DR>0.8$. Events where the number of AK4 jets passing the loose \textsc{DeepJet} WP (\Nbjets) is greater than zero are vetoed. These selection criteria are summarized in Table~\ref{sel:Bcat}, along with those of the CRs.

\begin{figure*}[!ht]
  \centering
  \includegraphics[width=0.49\textwidth]{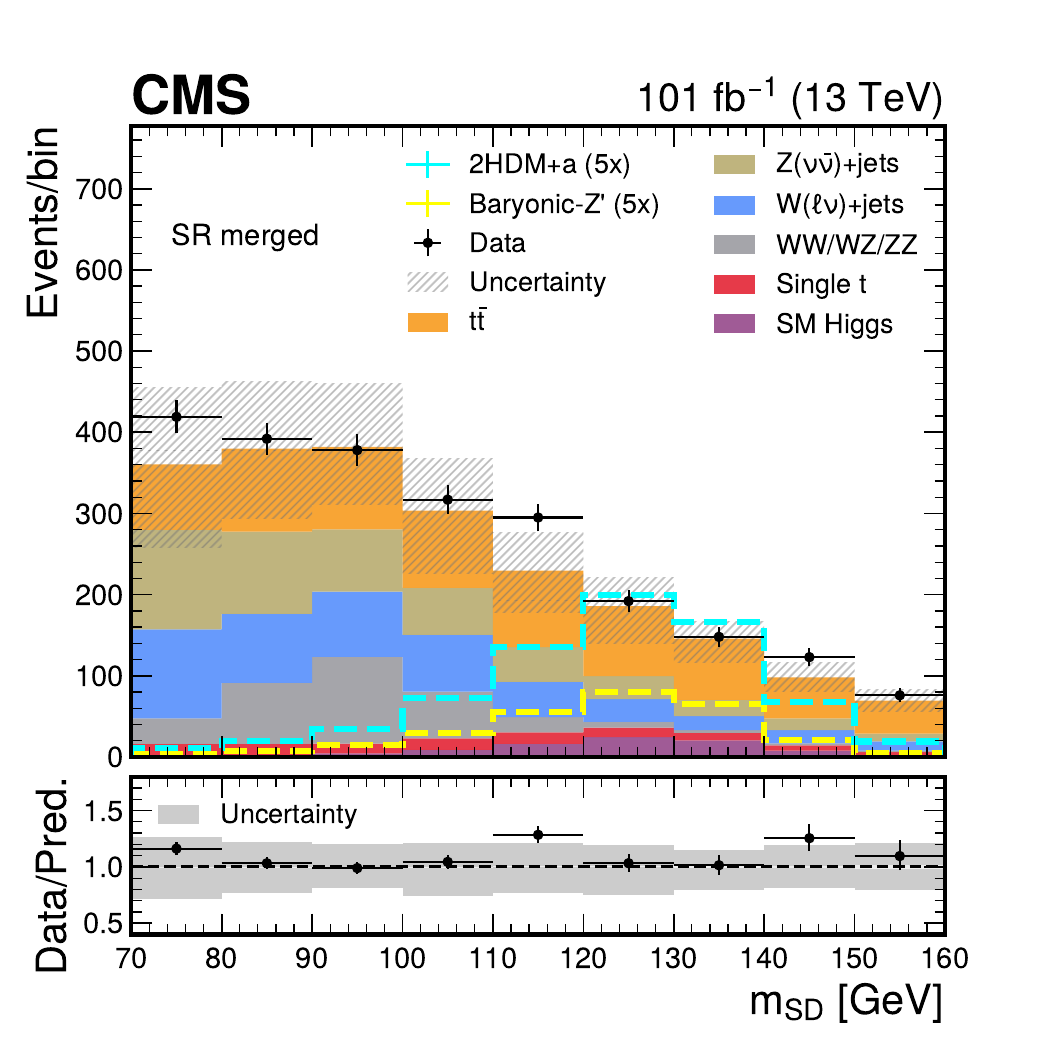} 
  \includegraphics[width=0.49\textwidth]{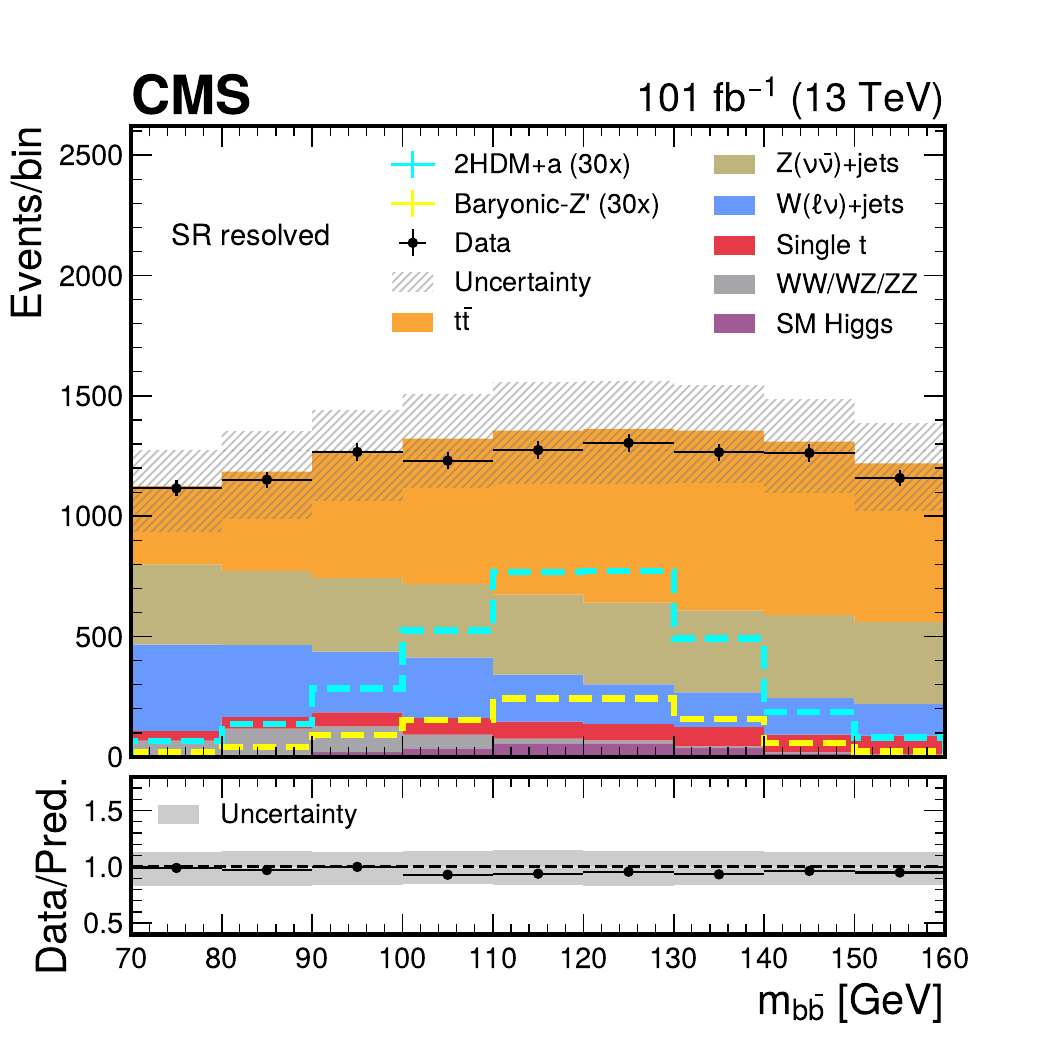}
  \caption{Distributions in the AK8 jet soft-drop mass $\mSD$ for the merged category (left), and in the dijet mass $\mbb$ for the resolved category (right), SRs. The black markers with statistical uncertainty bars show the observed data, the stacked colored histograms show the predicted pre-fit background, and the gray shading shows the statistical uncertainty in the prediction. The signal predictions, scaled up for visibility by the factors indicated in the legends, are overlaid as yellow and cyan dashed lines, one for each benchmark model: \baryonicZprime with $(\mZpr=2250, \mchi=1) \GeV$ and \twohdma with $(\mA=1000, \ma=350) \GeV$, respectively. The lower panels show the ratios of the observed data to the predicted background, with the gray bands indicating the systematic uncertainty.}
  \label{fig:HiggsCandidateMass}
\end{figure*}

In the resolved-category SR, the two \PQb-tagged AK4 jets must pass \pt thresholds of 50 and 30\GeV. To reduce the \ttbar background by a factor of two while preserving high acceptance for signal events, the \PSh candidate, \ie, the ``dijet'' system formed by the two \PQb-tagged AK4 jets, must have transverse momentum $\pt^{\smash[b]{\bbbar}}>100\GeV$ and invariant mass ($\mbb$) within 70--160\GeV. We impose a requirement similar to the one for the merged category to suppress the \ttbar background: $\Njets\leq2$, counting AK4 jets with $\pt>30\GeV$ within $\DR>0.4$ of either \PQb-tagged jet forming the \PSh candidate. These selection criteria are summarized in Table~\ref{sel:Rcat}, along with those of the CRs. Figure~\ref{fig:HiggsCandidateMass} shows the AK8 jet $\mSD$ distribution in the merged category (left) and the dijet $\mbb$ distribution in the resolved category (right).

The QCD multijet events can mimic the signal-like final state with a large artificial \ptmiss arising from jet momentum mismeasurements or from failures in the PF reconstruction. In the first scenario, the events have a characteristic feature with small azimuthal angle difference between the misreconstructed jet \ptvecjet and the \ptvecmiss vector. To ensure that a jet mismeasurement is not the cause of \ptmiss, the \Dphimissjet for AK4 jets is required to be larger than 0.4, which is the typical angular size of an AK4 jet. Additionally, in the merged category, requiring $\Dphimissakeight>0.8$ between \ptvecmiss and the AK8 jet, as well as in the resolved category, requiring $\Delta\phi(\ptvecmiss,\ptvecbbbar)>1.6$ between \ptvecmiss and the dijet system, reduces the QCD background contamination significantly. In the second scenario, the events in the SR are required to have a maximum azimuthal angular separation between the PF-based and charged-particle track-based \ptvecmiss of $\Dphipftrack<2$. The QCD multijet background passing these requirements is less than 1\% of the total background.

\subsection{Background estimation and control regions}

\begin{figure*}[!ht]
  \centering
  \includegraphics[width=\textwidth]{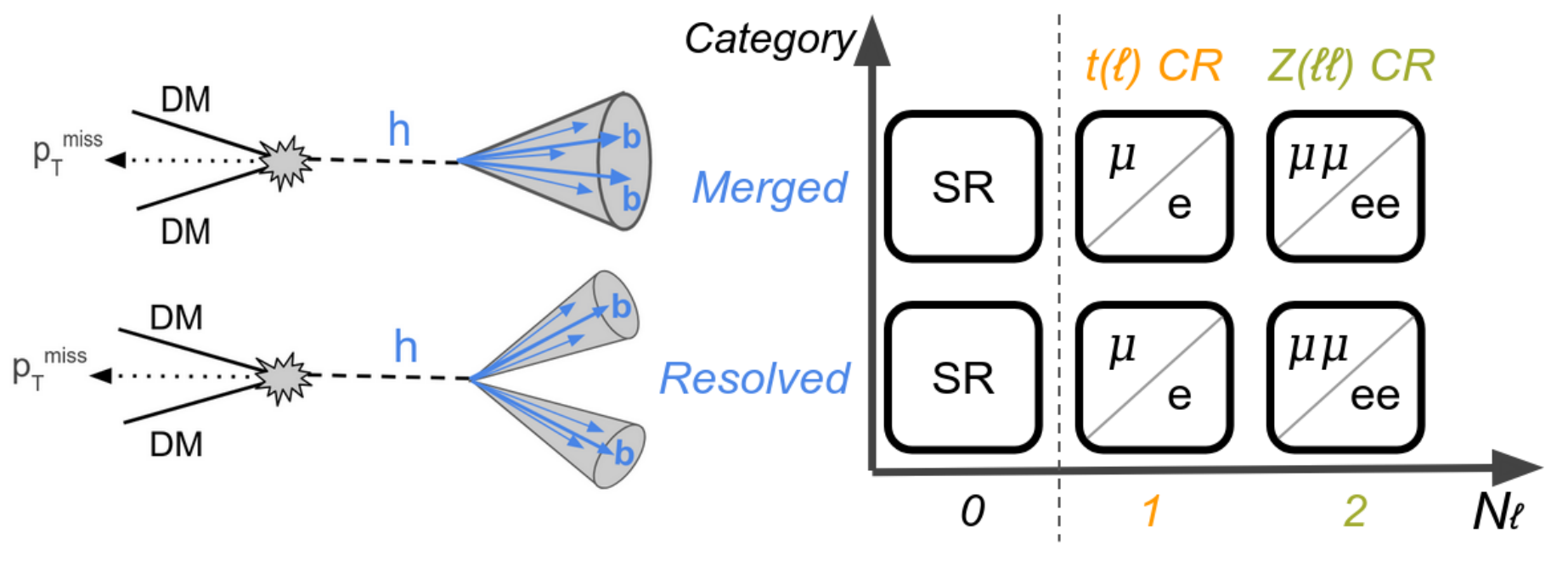}
  \caption{Schematic representation of the analysis regions.}
  \label{fig:schematic_regions}
\end{figure*}

The dominant background processes, \ttbar and \Znunujets, are estimated separately in each SR category using dedicated CRs, as detailed in Section~\ref{sec:signalExtraction}. Single-lepton CRs are used to estimate the \ttbar contribution, of which the lepton plus jets decay is the most important, while dilepton CRs, enriched in \Zlljets events, are used to estimate the \Znunujets background. The CRs are built by requiring exactly one or two well-reconstructed electrons or muons that satisfy stringent quality criteria. These regions are illustrated in Fig.~\ref{fig:schematic_regions} with single-lepton CRs---referred to as \topl CRs---and dilepton CRs---referred to as \zll CRs---defined separately for each SR category.

Events in the electron CRs are selected using a single-electron trigger, with at least one electron with $\pt>32\GeV$. Events in the muon CRs are selected using the \ptmiss trigger with a \ptmiss threshold of 120\GeV, as discussed in Section~\ref{subsection:SR}.

The hadronic recoil \vecrecoil is defined by removing the electron(s) or muon(s) \ptvecl from the \ptvecmiss calculation in the CRs, and its magnitude \recoil acts as a proxy for \ptmiss in the SRs. Events in the CRs are required to have $\recoil>250\GeV$ in the merged and $\recoil>200\GeV$ in the resolved category.

In each of the CRs, a similar selection strategy as in the corresponding SR (merged or resolved) is imposed, except for the charged-lepton selection criterion. In the \topl CRs, exactly one electron (muon) with $\pt>35$ (30)\GeV is required that passes tight identification and isolation quality criteria, corresponding to an identification efficiency of 90 (95)\%. In the \zll CRs, two oppositely charged electrons or muons are required that satisfy the same set of tight quality criteria. Furthermore, the dilepton \pt must be above 200\GeV with the dilepton invariant mass in the \PZ boson mass window, $60<m_{\Pell\Pell}<120\GeV$.

To accurately model the \Znunujets background process in the SRs, a sufficient number of \Zlljets events are retained in the \zll CRs by relaxing the \PQb-tagging condition for the \PSh candidate. In the merged category, the \textsc{ParticleNet-MD} double-\PQb tagging requirement for the AK8 jet is dropped. Similarly, in the resolved category, we drop the \textsc{DeepJet} \PQb tagging requirement for the AK4 jet candidates making up the \PSh candidate dijet system. 

In the \topl CR of the merged category, one of the \PQb jets in \ttbar events is captured by the AK8 jet identified with the \textsc{ParticleNet-MD} double-\PQb algorithm. To capture the second \PQb jet in a \ttbar event, we require an additional \PQb-tagged AK4 jet that is separated from the AK8 jet by $\DR>0.8$. In the \topl CR of the resolved category, the presence of two AK4 \PQb tagged jets is already accounted for.

Furthermore, in the \topl CRs, events are required to satisfy $\ptmiss>50\GeV$ to select events with the lepton+jets decay of the \ttbar pair while suppressing other background contributions. For the \zll CRs, $\ptmiss<100\GeV$ is required to ensure a high purity of \Zlljets events and to remain in a kinematic region where the modeling of \ptmiss in simulation is well understood. Tables~\ref{sel:Bcat} and \ref{sel:Rcat} summarize the merged and resolved event selections in the respective categories for the SRs and CRs.

\begin{table*}[!ht]
\centering
\topcaption{Event selections applied to the merged-category SR and CRs. Events in all the analysis regions have a photon and tau lepton candidate veto. Here ``jets'' refers to the AK4 jets with $\DR>0.8$ between the jet and the double-\PQb tagged AK8 jet.}
\label{sel:Bcat}
\renewcommand{\arraystretch}{1.2}
\cmsTable{\begin{scotch}{lccc}
    Variable & \multicolumn{3}{c}{Merged category selection} \\
    & SR & \topl CR & \zll CR \\  \hline
    \ptmiss ({\GeVns}) & $>$250 & $>$50 & \NA \\
    \recoil ({\GeVns}) & \NA & $>$250 & $>$250 \\
    $N_{\text{AK8 jets}}$ & \renewcommand{\arraystretch}{1}\begin{tabular}[t]{@{}c@{}} 1 \\{} [$\pt>200\GeV$, \\$70<\mSD<160\GeV$,\\ double-\PQb tagged] \end{tabular} & \renewcommand{\arraystretch}{1}\begin{tabular}[t]{@{}c@{}} 1 \\{} [$\pt>200\GeV$, \\$70<\mSD<160\GeV$,\\ double-\PQb tagged] \end{tabular} & \renewcommand{\arraystretch}{1}\begin{tabular}[t]{@{}c@{}} 1 \\{} [$\pt>200\GeV$, \\$70<\mSD<160\GeV$] \end{tabular} \\
    \Njets & $\leq$2 & $\leq$2 & $\leq$2 \\
    \Nbjets & 0 & 1 & 0 \\
    \Nell & 0 & 1 & 2 \\
    $\pt^{\Pell\Pell}$ ({\GeVns}) & \NA & \NA & $>$200 \\
    $m_{\Pell\Pell}$ ({\GeVns}) & \NA & \NA & $\in$[60--120] \\
    \Dphimissjet & $>$0.4 & \NA & \NA \\
    \Dphimissakeight & $>$0.8 & \NA & \NA \\
    \Dphipftrack & $<$2.0 & \NA & \NA \\
\end{scotch}}
\end{table*}

\begin{table}[!ht]
\topcaption{Event selections applied to the resolved-category SR and CRs. Events in all the analysis regions have a photon and tau lepton candidate veto. Here ``jets'' refers to the AK4 jets with $\DR>0.4$ between this jet and the leading ($\PQj_1$) and subleading ($\PQj_2$) \PQb-tagged jets forming the Higgs boson candidate dijet ($\PQj\PQj$) system.}
\label{sel:Rcat}
\centering
\renewcommand{\arraystretch}{1.2}
\begin{scotch}{lccc}
    Variable & \multicolumn{3}{c}{Resolved category selection} \\
    & SR & \topl CR & \zll CR \\ \hline
    \ptmiss ({\GeVns}) & $>$200 & $>$50 & $<$100 \\
    \recoil ({\GeVns}) & \NA & $>$200 & $>$200 \\
    $\pt^{\PQj_1}$, $\pt^{\PQj_2}$ ({\GeVns}) & $>$50, $>$30 & $>$50, $>$30 & $>$50, $>$30 \\
    & \PQb-tagged & \PQb-tagged & \NA \\
    $\pt^{\PQj\PQj}$ ({\GeVns}) & $>$100 & $>$100 & $>$100 \\
    $m_{\PQj\PQj}$ ({\GeVns}) & $\in$[70--160] & $\in$[70--160] & $\in$[70--160] \\
    \Njets & $\leq$2 & $\geq$1 & $\leq$2 \\
    \Nell & 0 & 1 & 2 \\
    $\pt^{\Pell\Pell}$ ({\GeVns}) & \NA & \NA & $>$200 \\
    $m_{\Pell\Pell}$ ({\GeVns}) & \NA & \NA & $\in$[60--120] \\
    \Dphimissjet & $>$0.4 & \NA & \NA \\
    $\Delta\phi(\ptvecmiss,\ptvec^{\,\PQj\PQj})$ & $>$1.6 & \NA & \NA \\
    \Dphipftrack & $<$2.0 & \NA & \NA \\
\end{scotch}
\end{table}

\subsection{Signal extraction}
\label{sec:signalExtraction}

The analysis is performed in the merged and resolved categories using a two-dimensional maximum likelihood fit. The observables are \ptmiss and the Higgs boson candidate mass \mSD in the merged category and \mbb in the resolved category. The Higgs boson candidate mass provides strong discrimination, as only the signal and well-understood, subdominant SM backgrounds exhibit a resonant peak in this distribution. The signal hypotheses typically exhibit harder \ptmiss than the dominant SM backgrounds, leading to a progressively improved signal-to-background ratio at high \ptmiss. To enhance the sensitivity of the analysis, we slice into exclusive \ptmiss regions and perform a simultaneous fit across all regions. A variable binning is used for \ptmiss in SRs (\recoil in CRs), which is optimized for signal sensitivity and tailored separately for each category. The bin edges are 250, 350, and 500\GeV for the merged category, and 200, 250, 290, 360, and 420\GeV for the resolved category. Thus, the signal is extracted using both the SRs and the corresponding CRs simultaneously in the binned maximum likelihood fit.

For the dominant backgrounds, \ttbar and \Znunujets, the Higgs boson candidate mass distribution in each \ptmiss window of the SRs is tied to the corresponding \recoil window in the CRs via a scale factor whose systematic uncertainty is accounted for with a nuisance parameter in the fit. For each category, the electron and muon CRs are combined appropriately to form a single CR for each background estimated from data. The contributions from other background processes are taken directly from MC simulation, with their nominal expected yield obtained after applying all event selections and normalizing them to the integrated luminosity times their respective theoretical cross sections.

The fit to the data determines the background post-fit yields in each bin, which are obtained as the pre-fit yields adjusted by the post-fit values of the scale factors through their associated nuisance parameters. The systematic uncertainties, many of which are shape based, account for experimental sources, such as jet energy scale and resolution, \PQb tagging, lepton identification and trigger efficiencies, as well as theoretical uncertainties affecting the background and signal distributions, detailed in Section~\ref{sec:Syst}. The shape-based systematic uncertainties further ensure that any residual mismodeling is accounted for in the total uncertainty.

The final free parameter in the fit is the signal strength modifier $\mu$, which controls the signal normalization relative to the theoretical cross section for a given signal hypothesis. The best fit background model, and the best fit signal strength are simultaneously obtained by maximizing the likelihood fit across the two categories, all SRs and CRs, and all data-taking periods.

\begin{figure*}[!htp]
  \centering
  \includegraphics[width=0.8\textwidth]{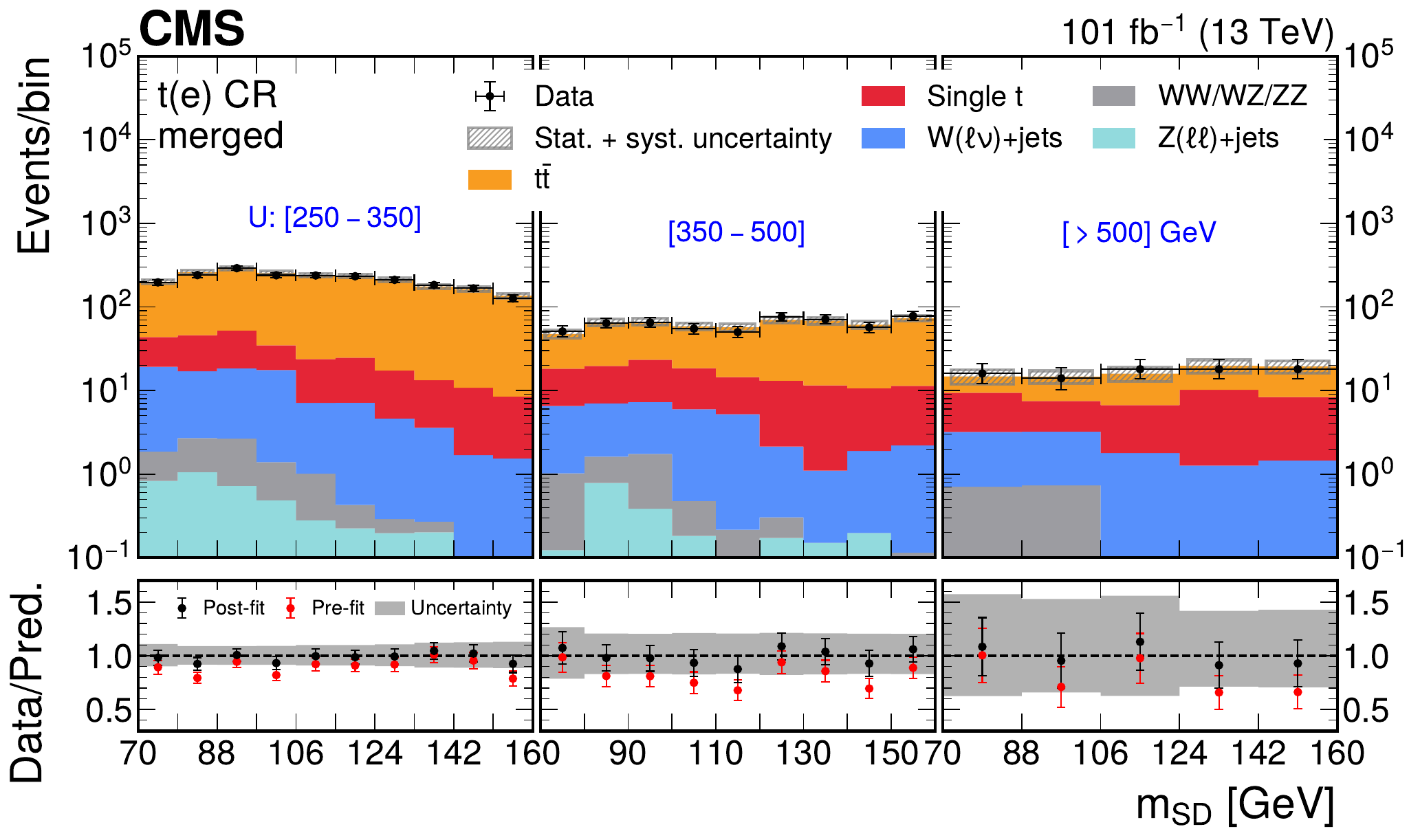} \\
  \includegraphics[width=0.8\textwidth]{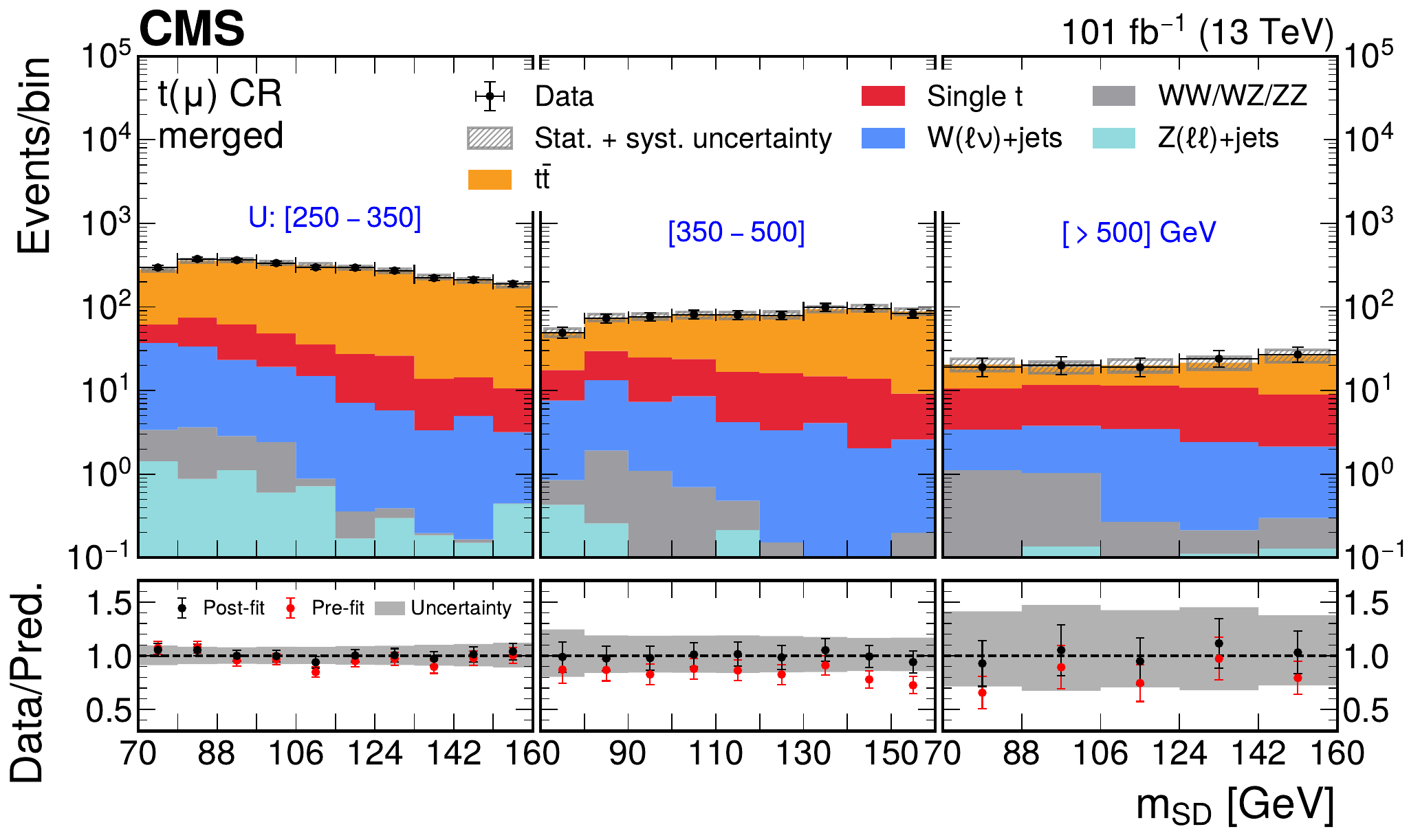}
\caption{The AK8 jet \mSD distributions after the simultaneous likelihood background-only fit sliced in three \recoil bins, for the merged-category $\PQt(\Pe)$ (above) and $\PQt(\PGm)$ (below) CRs. The black markers with statistical uncertainty bars show the observed data, the stacked colored histograms show the predicted background, and the gray shading shows the systematic uncertainty in the prediction. The lower panels show the ratios of the observed data to the background predicted before (pre-fit in red) and after (post-fit in black) the fit, with the gray bands indicating the post-fit uncertainty obtained from combining all statistical and systematic sources.}
\label{postfitTopCRs:B_1718}
\end{figure*}

\begin{figure*}[!htp]
  \centering
  \includegraphics[width=0.8\textwidth]{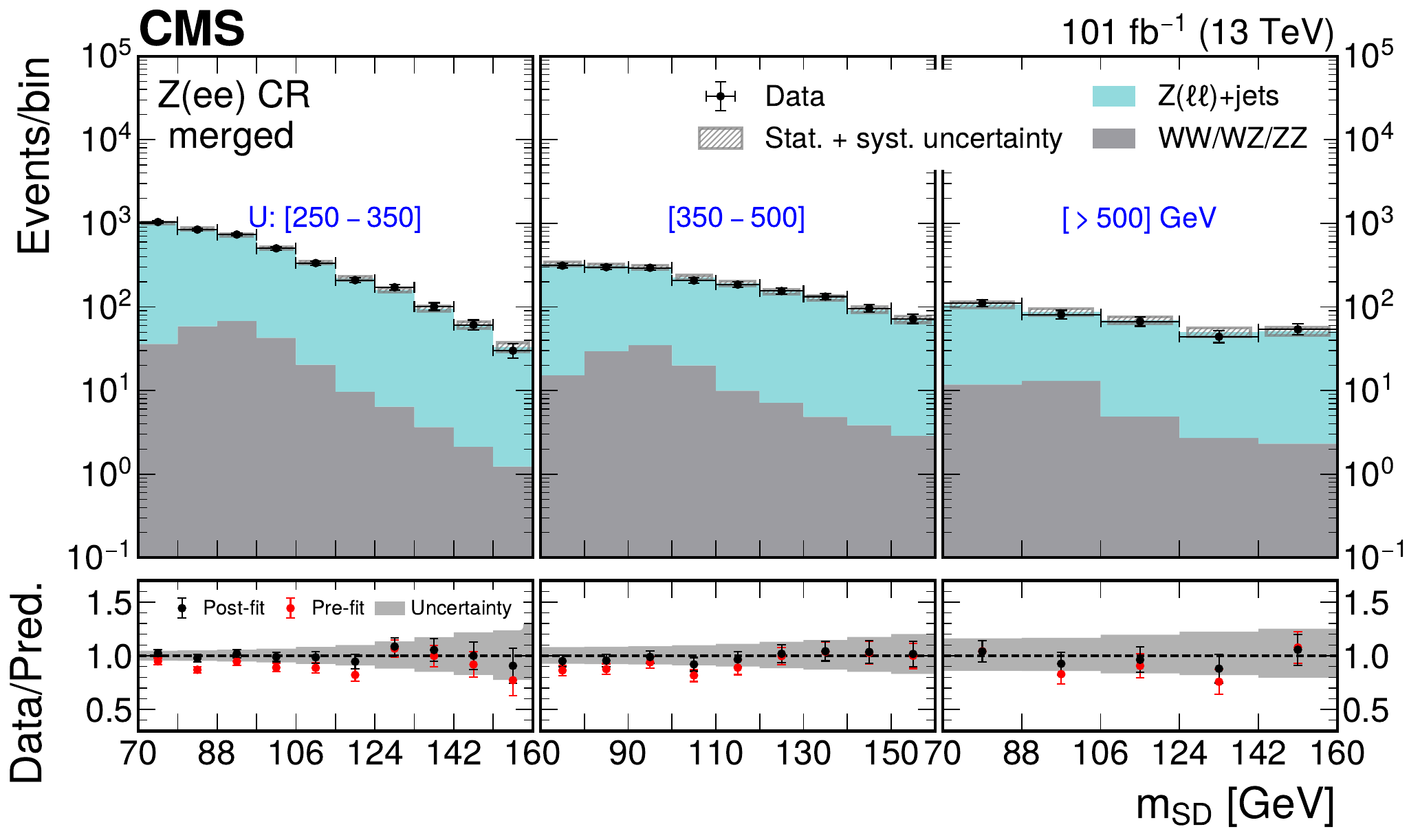} \\
  \includegraphics[width=0.8\textwidth]{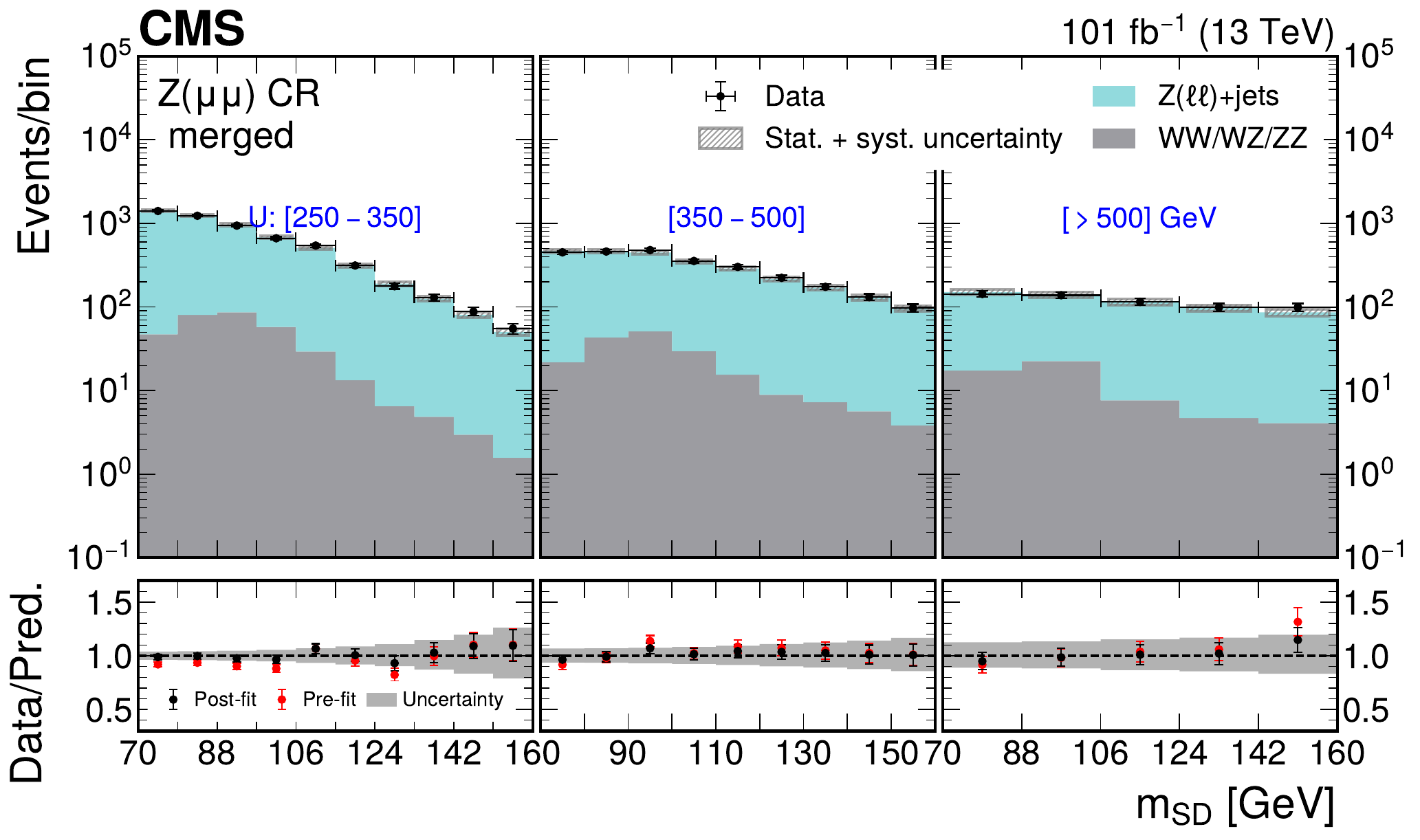}
\caption{The AK8 jet \mSD distributions after the simultaneous likelihood background-only fit sliced in three \recoil bins, for the merged-category $\PZ(\Pe\Pe)$ (above) and $\PZ(\PGm\PGm)$ (below) CRs. The black markers with statistical uncertainty bars show the observed data, the stacked colored histograms show the predicted background, and the gray shading shows the systematic uncertainty in the prediction. The lower panels show the ratios of the observed data to the background predicted before (pre-fit in red) and after (post-fit in black) the fit, with the gray bands indicating the post-fit uncertainty obtained from combining all statistical and systematic sources.}
\label{postfitZCRs:B_1718}
\end{figure*}

\begin{figure*}[!htp]
  \centering
  \includegraphics[width=\textwidth]{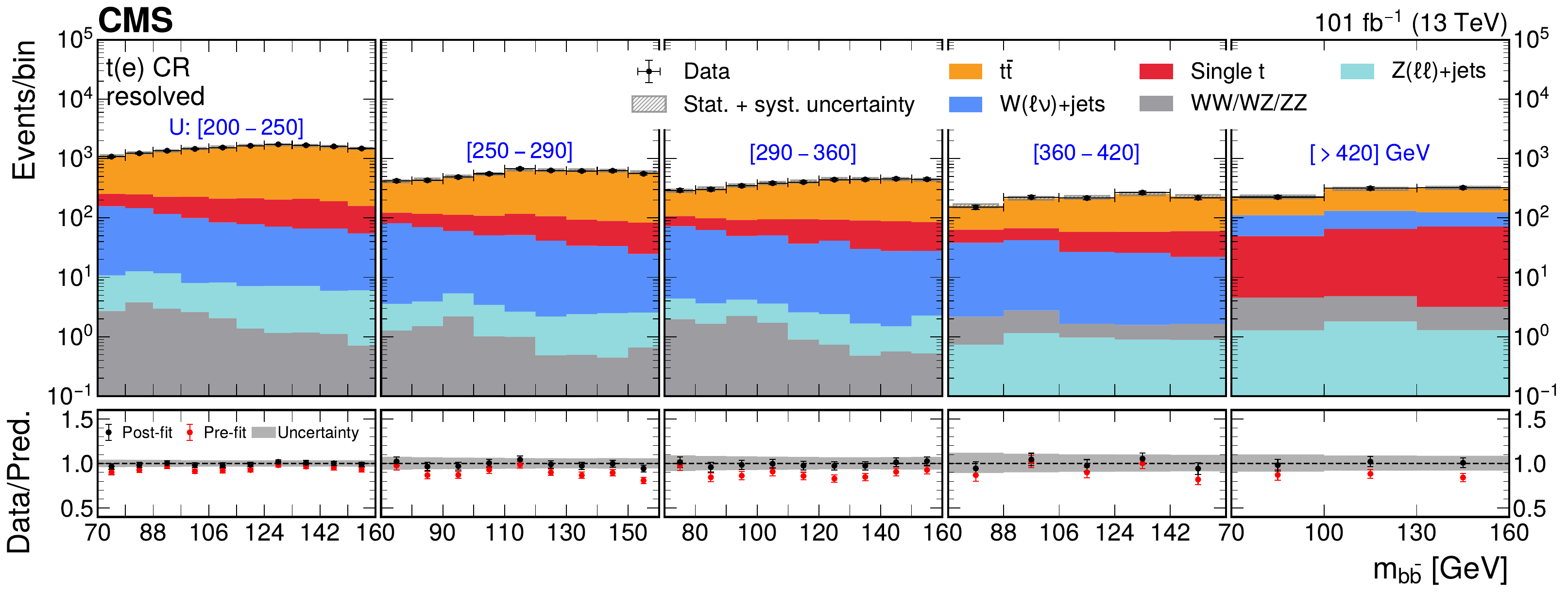} \\
  \includegraphics[width=\textwidth]{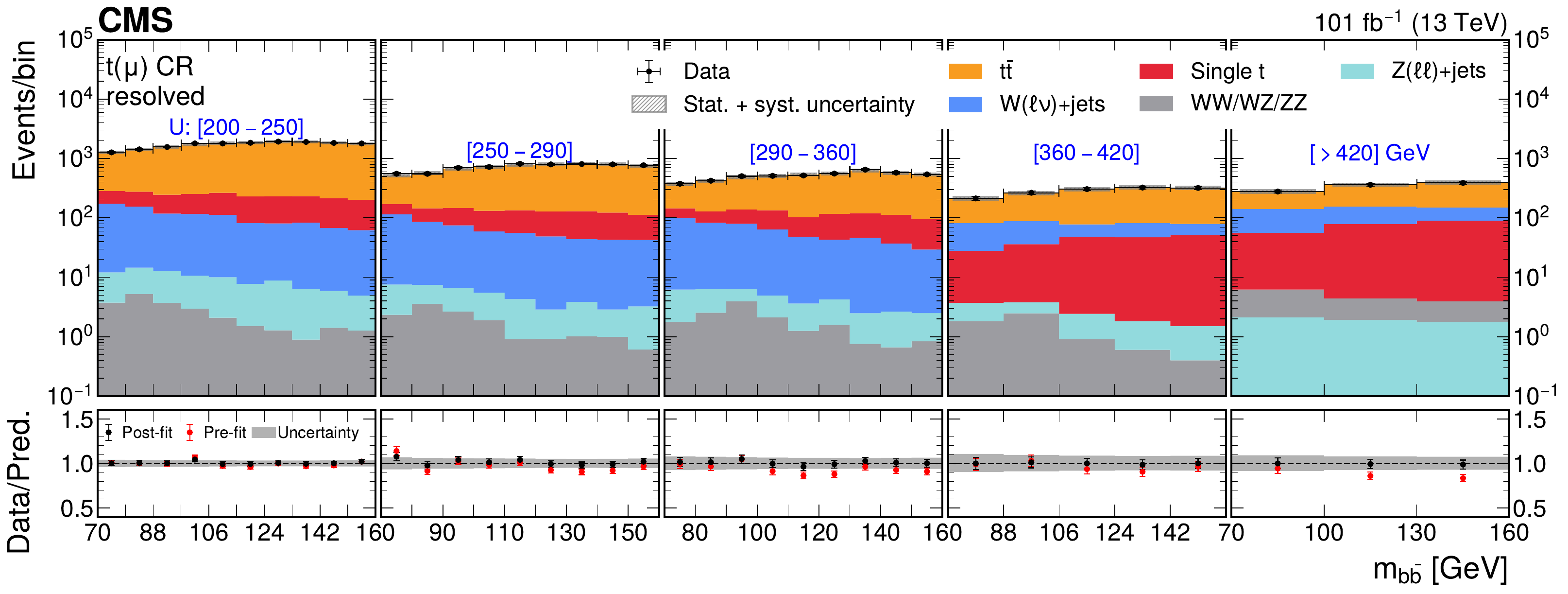}
\caption{The dijet \mbb distributions after the simultaneous likelihood background-only fit sliced in five \recoil bins, for the resolved-category $\PQt(\Pe)$ (above) and $\PQt(\PGm)$ (below) CRs. The black markers with statistical uncertainty bars show the observed data, the stacked colored histograms show the predicted background, and the gray shading shows the systematic uncertainty in the prediction. The lower panels show the ratios of the observed data to the background predicted before (pre-fit in red) and after (post-fit in black) the fit, with the gray bands indicating the post-fit uncertainty obtained from combining all statistical and systematic sources.}
\label{postfitTopCRs:R_1718}
\end{figure*}

\begin{figure*}[!htp]
  \centering
  \includegraphics[width=\textwidth]{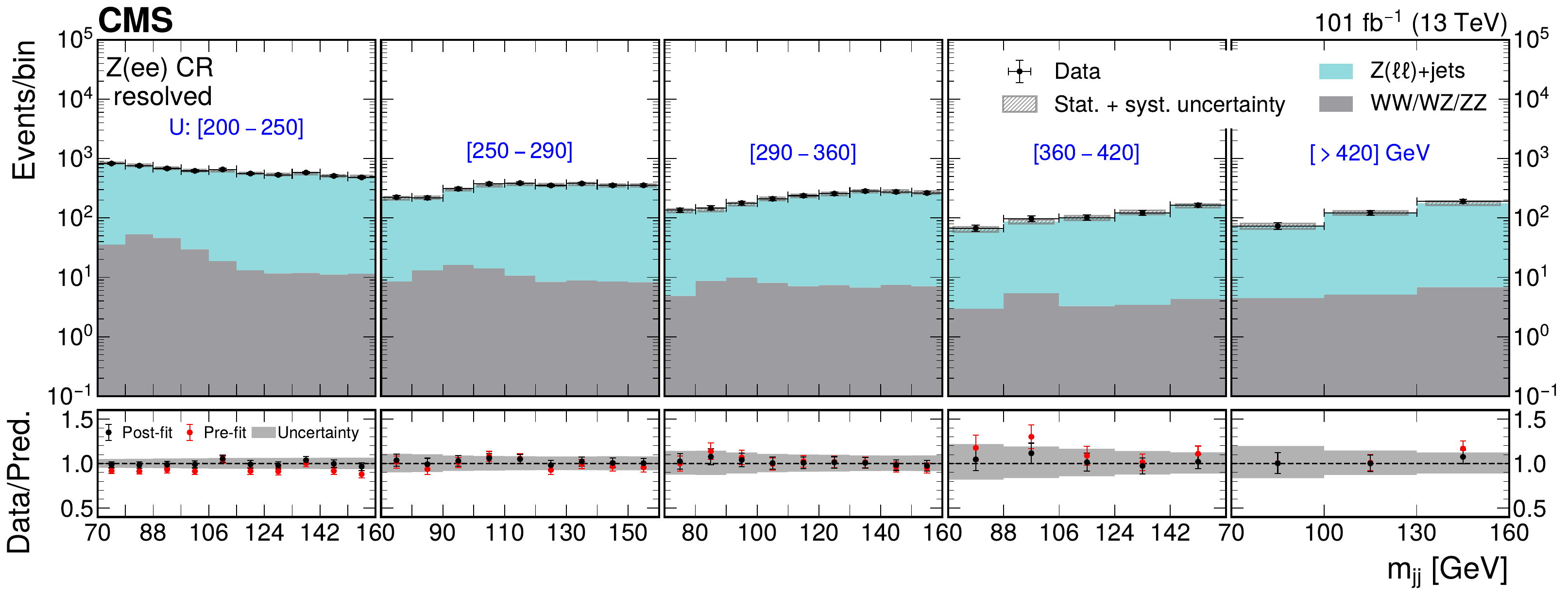} \\
  \includegraphics[width=\textwidth]{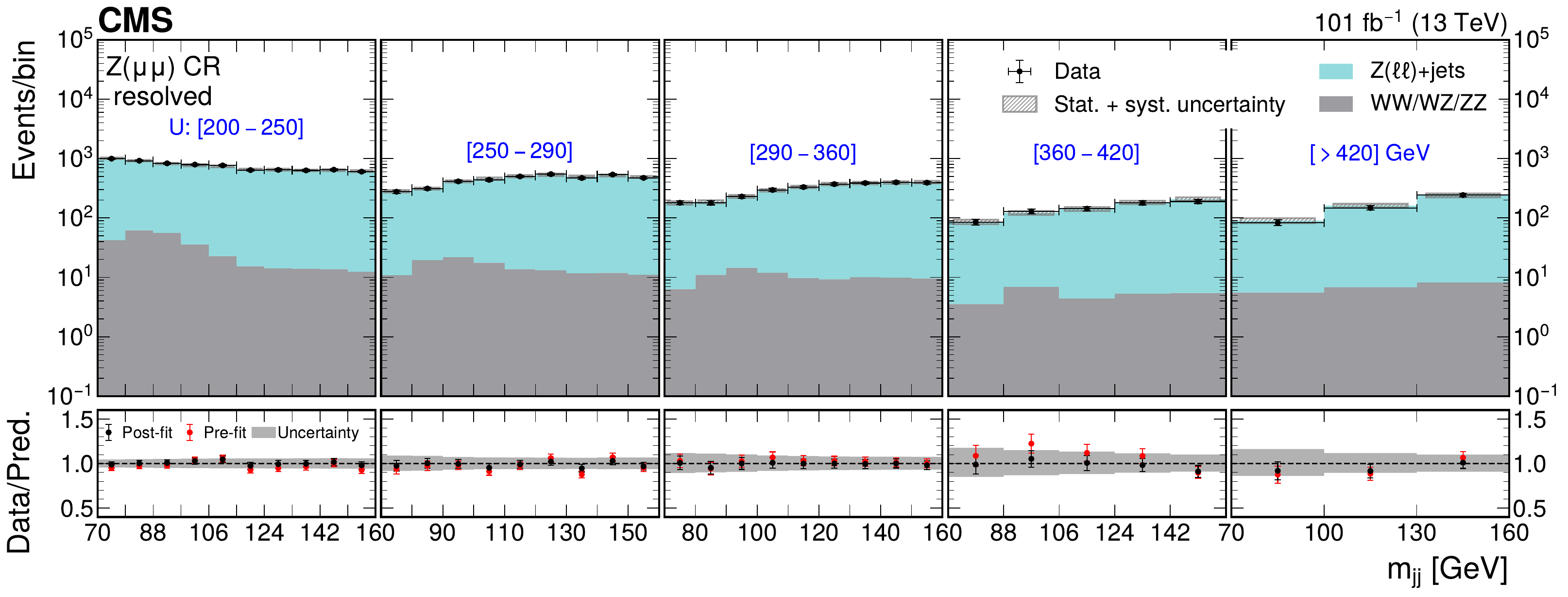}
\caption{The dijet $m_{\PQj\PQj}$ distributions after the simultaneous likelihood background-only fit sliced in five \recoil bins, for the resolved-category $\PZ(\Pe\Pe)$ (above) and $\PZ(\PGm\PGm)$ (below) CRs. The black markers with statistical uncertainty bars show the observed data, the stacked colored histograms show the predicted background, and the gray shading shows the systematic uncertainty in the prediction. The lower panels show the ratios of the observed data to the background predicted before (pre-fit in red) and after (post-fit in black) the fit, with the gray bands indicating the post-fit uncertainty obtained from combining all statistical and systematic sources.}
\label{postfitZCRs:R_1718}
\end{figure*}

Figures~\ref{postfitTopCRs:B_1718} and \ref{postfitZCRs:B_1718} show the post-fit \PSh candidate mass distributions in the merged-category \topl and \zll CRs, respectively, and Figs.~\ref{postfitTopCRs:R_1718} and \ref{postfitZCRs:R_1718} show the corresponding distributions for the resolved-category CRs. The lower panels display the ratio of the yields in data to the prediction before (pre-fit) and after (post-fit) the fit. The agreement of these post-fit yields with those observed in the data in the CRs provides additional confidence in the method.

\section{Systematic uncertainties}
\label{sec:Syst}

All systematic uncertainties affecting the simulated processes (signal and background) in the analysis regions are treated as nuisance parameters within the likelihood function. These parameters influence either the normalization or the shape of the \ptmiss (\recoil) distribution for a given process in the SR (CR) and are constrained in the simultaneous fit.

Uncertainties that do not affect the shape of the distribution, or are uniformly distributed are treated as normalization uncertainties and incorporated using log-normal distributions. Those having nonuniform variations are treated as shape uncertainties and are incorporated by means of Gaussian prior distributions. Experimental and theoretical sources of systematic uncertainty considered in this search are described below.

The \ptmiss trigger efficiency is measured using single-muon events passing the single-muon trigger with a \pt threshold of 27 (24)\GeV for data recorded in 2017 (2018). It is parametrized as a function of the \recoil variable, and the corresponding data-to-simulation correction factors are applied to the simulation in the SRs and the single-muon and dimuon CRs. The electron trigger efficiency is parametrized as a function of the electron \pt and $\eta$. An associated uncertainty of 1--2\% is assigned to both the trigger scale factors.

Lepton reconstruction and identification uncertainties are evaluated and incorporated as a function of \pt and $\eta$. The systematic uncertainties related to muon identification and isolation have a normalization effect of less than 1\%. For electrons, the systematic uncertainties associated with reconstruction (identification) have a normalization effect of 1--2 (1--5)\%. An uncertainty of 6 (2)\% is assigned to the tau lepton (photon) veto efficiency across all analysis regions.

The integrated luminosities for the 2016, 2017, and 2018 data-taking years have 1.2--2.5\% individual uncertainties~\cite{CMS-LUM-17-003,CMS-PAS-LUM-17-004,CMS-PAS-LUM-18-002,CMS-PAS-LUM-20-001}, while the overall uncertainty for the 2016--2018 period is 1.6\%. A normalization uncertainty of 2.5\% is applied to all background processes taken directly from the MC simulations.

Pileup modeling is addressed by reweighting the signal and background MC samples such that the number of true pileup interactions in simulation matches the pileup profile observed in data. The latter is determined using integrated luminosity per bunch crossing and assuming the total inelastic cross section of 69.2\unit{mb}~\cite{CMS:2018mlc}. The systematic uncertainty from the pileup reweighting is evaluated by varying the total inelastic cross section by $\pm$4.6\%.

During the 2017 data-taking period, a gradual shift in the timing of the inputs of the ECAL L1 trigger in the $\abs{\eta}>2.0$ region caused a trigger inefficiency, denoted as ``prefiring''~\cite{CMS:2020cmk}. The trigger missed between 10 and 20\% of the events containing an electron (of $\pt\gtrsim50\GeV$) or a jet (of $\pt\gtrsim100\GeV$) in the $2.5<\abs{\eta}<3.0$ range, the exact loss depending on the \pt and $\eta$ of the trigger object. Correction factors were computed from data and applied to the acceptance evaluated by simulation. The shape-based uncertainty for L1 prefiring has a normalization effect of 0.5--1\%.

The JES corrections depend on jet \pt and $\eta$, and the uncertainties are split into 11 independent sources accounting for different experimental effects~\cite{CMS:2016lmd}. These systematic uncertainties are considered for all simulation-based processes, with a shape effect ranging from 0.1 to 5\%. Differences in the JER between data and simulation are accounted for by smearing the \pt of jets in MC samples. The smearing factors and their uncertainties depend on the jet $\eta$. The shape-based uncertainty is assigned to all simulated processes, with an effect of around 4\%. The change in the jet momentum due to JER smearing is propagated to the \ptmiss computation.

The efficiency of the \textsc{DeepJet}~\cite{Bols_2020, CMS-DP-2023-005} \PQb tagging algorithm is parametrized as a function of jet \pt, $\eta$, and flavor, and is applied to AK4 \PQb jets making the \PSh candidate in the resolved category. Discrepancies between data and simulation are corrected for using data-to-simulation correction factors. Shape-based uncertainties are computed separately for genuine \PQb jets and misidentified \PQb jets, with an average shape effect of 5\%. For the merged category, an additional uncertainty in the \textsc{ParticleNet-MD} double-\PQb tagging efficiency is applied to AK8 jets as a function of their \pt, with an average shape effect of around 2\%~\cite{PhysRevD.101.056019, CMS-PAS-BTV-22-001}.

Most experimental uncertainties are specific to each data set and are thus assumed to be uncorrelated between different data-taking periods. The uncertainties related to integrated luminosity, JES and JER, as well as those related to \PQb tagging are partially correlated between the data-taking periods, which is implemented by splitting the total uncertainty into its correlated and uncorrelated components.

To mitigate discrepancies between the data and MC simulations in the \pt spectra of the top quarks in \ttbar events, a parametric function is used to reweight the simulation. The uncertainty associated with this correction is estimated by comparing weighted and unweighted results, leading to a shape effect of around 7\%. Shape uncertainties for the LO-to-NLO scale factors obtained for \Wjets, \Znunujets, and \Zlljets processes~\cite{monojet_Run2} are considered to account for the missing higher-order corrections with QCD and EW terms. These shape uncertainties range from 5 to 20\%, depending on the category, process, and boson \pt.

Uncertainties in the PDFs and those due to variations in the renormalization and factorization scales ($\mu_{\text{R}}$ and $\mu_{\text{F}}$) impact both the shape and cross section of simulated processes. The PDF uncertainty is evaluated by reweighting simulated events using the set of PDF replicas from the NNPDF3.1 PDFs, following the PDF4LHC recommendations~\cite{Butterworth_2016}. The envelope of variations is taken as the systematic uncertainty, with a typical effect of 3--6\% depending on the process and category. Scale uncertainties are estimated by individually varying the $\mu_{\text{R}}$ and $\mu_{\text{F}}$ scales up and down by a factor of two, with the exception of simultaneously varying both up or down. The resulting uncertainty is 10--20\%, and is treated as uncorrelated for the signal as well as the various background processes. These uncertainties are treated as correlated across categories and data-taking eras.

For background processes taken directly from the simulations, a normalization uncertainty of 10\% is assigned to the single top quark process, 20\% to the \Wjets process, and 30\% to the diboson process. These values are chosen based on existing experimental constraints~\cite{CMS:2014fut,CMS:2010svw,CMS:2016ogx,CMS:2021icx}. For the signal process, a normalization uncertainty of 10\% is assigned. A cross section uncertainty of 5\% is assigned to the SM Higgs process. These uncertainties are correlated across categories and data-taking eras.

Table~\ref{tab:systematics} summarizes the sources of systematic uncertainties considered in the analysis, the correlation scheme between the 2017 and 2018 data-taking periods, along with the type (normalization or shape) and the relative effect on the \PSh candidate mass distribution sliced in \ptmiss windows.

\begin{table*}[t!]
\centering
\topcaption{The sources of systematic uncertainty and the correlation scheme between the 2017 and 2018 data-taking periods, along with the type (normalization or shape) and their relative values.}
\label{tab:systematics}
\renewcommand{\arraystretch}{1.1}
\cmsTable{\begin{scotch}{lccc}
    Systematic source & Correlation between years & Type & Relative uncertainty [\%]  \\
    \hline

    \PQb jet tagging efficiency & partially correlated  & shape & 1--5   \\
    \PQb jet mistagging efficiency & partially correlated  & shape & 1--4  \\

    Double-\PQb jet tagging efficiency & uncorrelated & shape & 0--2  \\

    Jet energy scale & partially correlated  & shape & 0.1--5  \\
    Jet energy resolution & uncorrelated  & shape & 0--4  \\

    \ptmiss trigger & uncorrelated & shape & 1--2  \\
    Electron trigger & uncorrelated & shape & 1--2  \\

    Electron reconstruction & uncorrelated & shape & 1--2  \\
    Electron identification & uncorrelated & shape & 1--5  \\
    Muon identification & uncorrelated & shape & ${<}$1  \\
    Muon isolation & uncorrelated & shape & ${<}$1  \\

    Integrated luminosity & partially correlated & normalization & 1.2--2.5 \\

    Pileup reweighting & uncorrelated & shape & 0--5  \\
    L1 prefiring & uncorrelated & shape & 0.5--1  \\

    Tau lepton veto & uncorrelated & normalization & 6  \\
    Photon veto & uncorrelated & normalization & 2  \\

    Top quark \pt reweight & correlated & shape & 0--7  \\
    \vjet \pt reweight & correlated & shape & 5--20  \\

    PDF & correlated  & shape & 3--6  \\
    Renorm./fact. scale & correlated  & shape & 0--20 \\

    Cross section single top quark & correlated & normalization & 10  \\
    Cross section \Wjets & correlated & normalization & 20  \\
    Cross section diboson & correlated & normalization & 30  \\
    Cross section Higgs boson & correlated & normalization & 5  \\
    Cross section signal & correlated & normalization & 10  \\
\end{scotch}}
\end{table*}

The MC bin-by-bin statistical uncertainties are handled using the CMS statistical tool \textsc{Combine}~\cite{Combine2024}, which employs the ``Barlow-Beeston-lite'' technique~\cite{BARLOW1993219}. A Poisson threshold of 10 events is set, such that for MC event counts exceeding 10, the uncertainty is modeled with a Gaussian distribution, while for counts below 10, a Poisson distribution is used to model per-process uncertainties in that bin.

The results based on the data set analyzed in this paper are further combined with the results from an earlier analysis based on data collected in 2016. The combination is performed by running a combined likelihood fit across all the analysis regions in all data sets. For this purpose, the uncertainty in integrated luminosity is partially correlated, and the PDF uncertainty is fully correlated between data-taking eras. The remaining uncertainty sources are treated as uncorrelated in the final combined fit.

\section{Results}
\label{sec:Result}

\begin{figure*}[!t]
  \centering
  \includegraphics[width=0.7\textwidth]{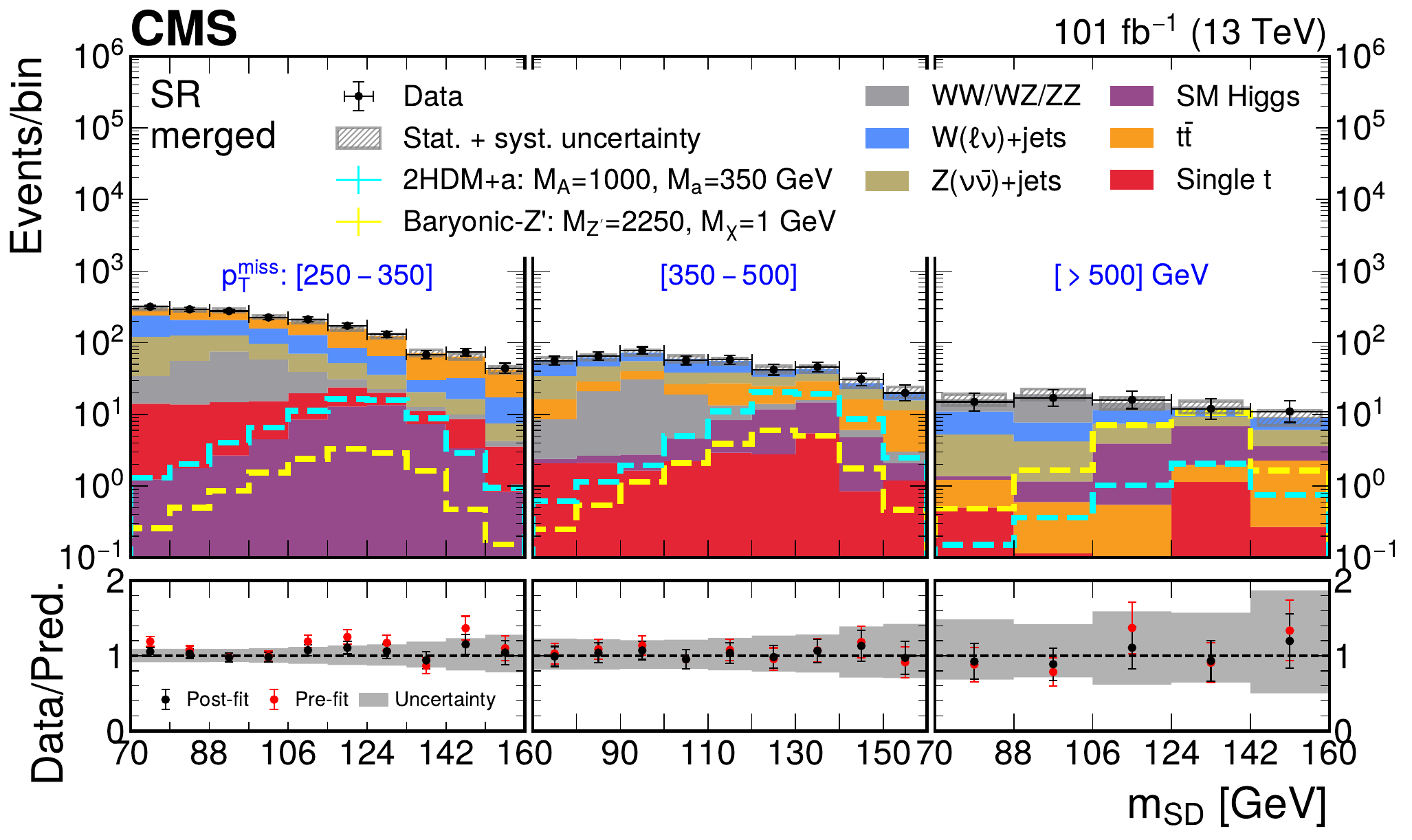} \\
  \includegraphics[width=\textwidth]{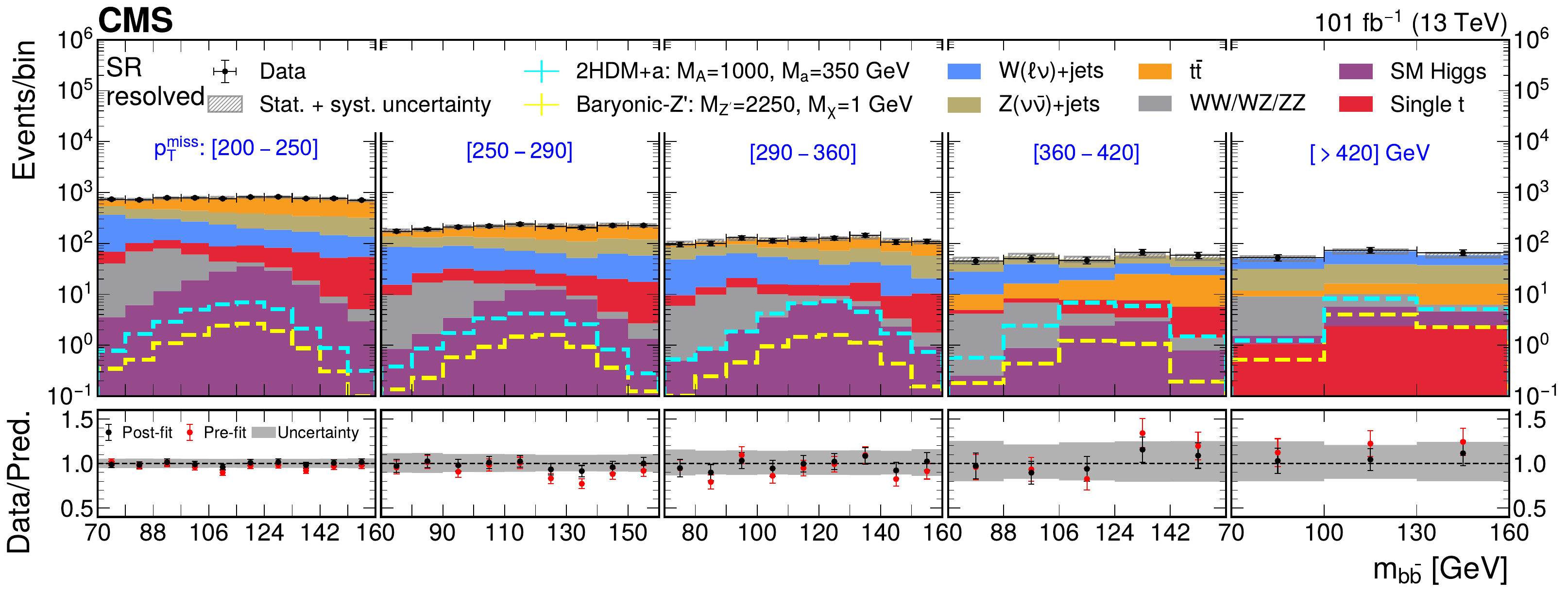}
  \caption{The AK8 jet \mSD (above) and dijet \mbb (below) distributions after the simultaneous likelihood background-only fit sliced in three and five \ptmiss bins, for the merged- and resolved-category SRs respectively. The black markers with statistical uncertainty bars show the observed data, the stacked colored histograms show the predicted background, and the gray shading shows the systematic uncertainty in the prediction. The signal predictions are overlaid as cyan and yellow dashed lines, one for each benchmark model. The lower panels show the ratios of the observed data to pre-fit (red points) and post-fit (black points) background predictions, with the gray bands indicating the post-fit uncertainty obtained from combining all statistical and systematic sources.}
  \label{postfitSR:B_R_1718}
\end{figure*}

A maximum-likelihood fit~\cite{Combine2024} to extract the signal yield is performed to the SRs and CRs by combining the merged and resolved categories, as well as the 2017 and 2018 data sets. Figure~\ref{postfitSR:B_R_1718} shows the \PSh candidate mass distributions in the \mSD and \mbb observables for the merged and resolved SRs, respectively, sliced into the three and five \ptmiss bins. The improved ratio of signal to background is evident as \ptmiss is increased, indicating the importance of fitting these regions separately, which increases signal sensitivity in the high \ptmiss regions while retaining the low \ptmiss regions that provide stronger constraints on the nuisance parameters describing the background. Superimposed on the data are the stacked histograms representing the background-only fit. The observed data show good agreement with this fit, and no significant excess of events is observed.

The results of this search are interpreted in two benchmark simplified models. An upper limit is computed on the cross section for the associated production of a dark matter candidate with a Higgs boson, calculated at 95\% confidence level (\CL)~\cite{Combine2024, CLS1, CLS2} using a modified frequentist approach~\cite{CMS-NOTE-2011-005} and asymptotic approximation~\cite{Cowan_2011}.

In this method, background-only and signal-plus-background fits are performed for each signal model. The nuisance parameters are profiled in the signal fit, and the best-fit value of the signal strength modifier $\mu$ is extracted for each signal hypothesis. The expected limit is obtained from background-only pseudodata, while the observed limit is derived from the data. Deviations between the expected and observed limit can arise from statistical fluctuations and nuisance-parameter profiling even when the background-only hypothesis provides the best fit to the observed data. A statistical combination is performed with the previous CMS analysis~\cite{exo:monohbb2016} that uses 2016 data. The interpretations presented in the following two sections are derived from the combined 2016--2018 data, with an integrated luminosity of 138\fbinv. For the branching fraction of $\PSh\to\bbbar$ we take 0.582~\cite{PDG2024}.

\subsection{Baryonic-\texorpdfstring{\PZpr}{Z'} model interpretation}

The \baryonicZprime model provides a theoretical framework in which the DM candidate \PGc couples to SM quarks via a massive vector mediator, \PZpr. The dominant kinematic feature of this model is the presence of large \ptmiss recoiling against a merged \PSh candidate jet, which motivates the use of high-\pt \PSh boson tagging. The parameter space is characterized primarily by the mediator and DM masses.

\newcommand{\sighDM}{\ensuremath{\sigma_{\PSh+\text{DM}}}\xspace}

\begin{figure*}[!ht]
    \centering
    \includegraphics[width=\textwidth]{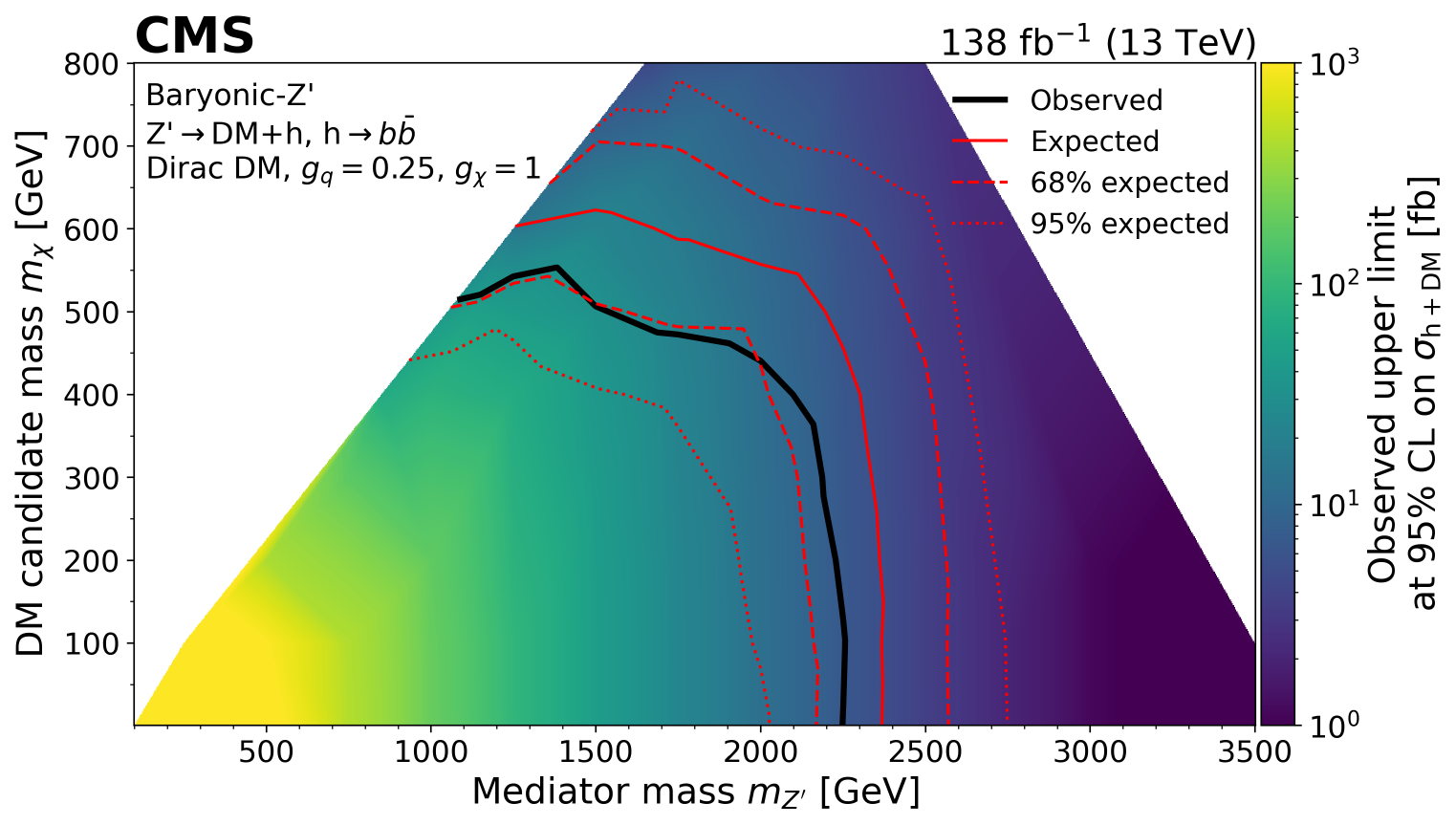}
    \caption{Exclusion limits at 95\% \CL on the signal cross section \sighDM for the \baryonicZprime model as a function of \mZpr and \mchi. The coupling parameters are fixed to $\gqua=0.25$ and $\gchi=1$. The areas within the solid black and red contours represent the exclusion regions where the theoretical cross sections are larger than the observed and expected experimental limits, respectively. The areas within the dashed and dotted red contours show the excluded regions at $\pm$1 and $\pm$2 standard deviations from the expected limits, respectively.}
    \label{limits:ZpB_2D}
\end{figure*}

Figure~\ref{limits:ZpB_2D} shows the exclusion limit at 95\% \CL on the signal cross section \sighDM, as a two-dimensional scan over \mZpr and \mchi. In this scan, the coupling between the mediator and SM quarks is fixed to $\gqua=0.25$, and the coupling between the mediator and DM particles is fixed to $\gchi=1$. For $\mchi=1\GeV$, values of \mZpr below 2.25\TeV (2.35\TeV expected) are excluded. Masses of DM particles up to 550\GeV (600\GeV expected) are excluded for $\mZpr=1.25\TeV$.

Compared to the previous result in Ref.~\cite{exo:monohbb2016} with the 2016 data set, the maximal exclusion limit in terms of \mZpr is improved by 650\GeV (500\GeV expected) for $\mchi=1\GeV$. The higher luminosity data contributes around 300\GeV of expected improvement, with the additional 200\GeV of expected exclusion coming from the improved identification of boosted $\PSh\to\bbbar$ decays with the latest tagging algorithms.

\subsection{The \texorpdfstring{\twohdma}{2HDM+a} model interpretation}
\label{subsec:2HDMa_results}

In the \twohdma model, the mono-Higgs signature arises from a scalar sector that includes a heavy pseudoscalar \PSA and a light pseudoscalar mediator \Pa, in addition to the \PSh boson. When the mass hierarchy $\mA>\ma+\mh$ is satisfied, the process $\PSA\to\Pa+\PSh$ becomes kinematically allowed, with the lighter pseudoscalar subsequently decaying invisibly to a pair of DM particles. The \twohdma framework offers a parameter space in which the production cross section and kinematic distributions are governed by model parameters such as the pseudoscalar masses \mA, \ma, the ratio of the vacuum expectation values \tanbeta, and the sine of the mixing angle, \sintheta. The DM particle is assumed to have a mass of $\mchi=10\GeV$. The masses of the heavy scalar, heavy pseudoscalar, and the two charged scalars are fixed to the same value, \ie, $\mH=\mA=\mHpm$~\cite{Bauer2017}.

\begin{figure*}[!ht]
  \centering
  \includegraphics[width=0.45\textwidth]{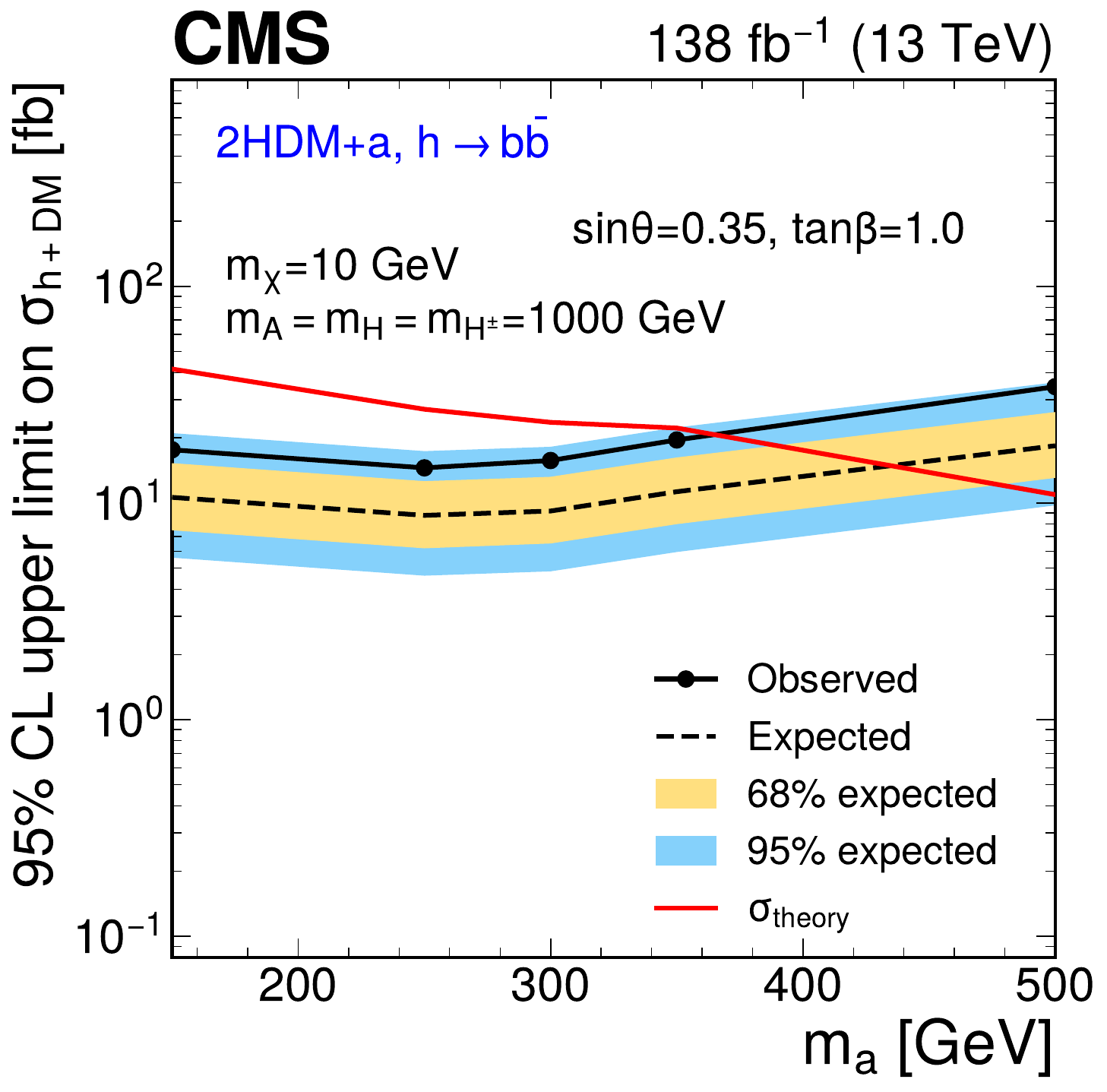}
  \hfill%
  \includegraphics[width=0.45\textwidth]{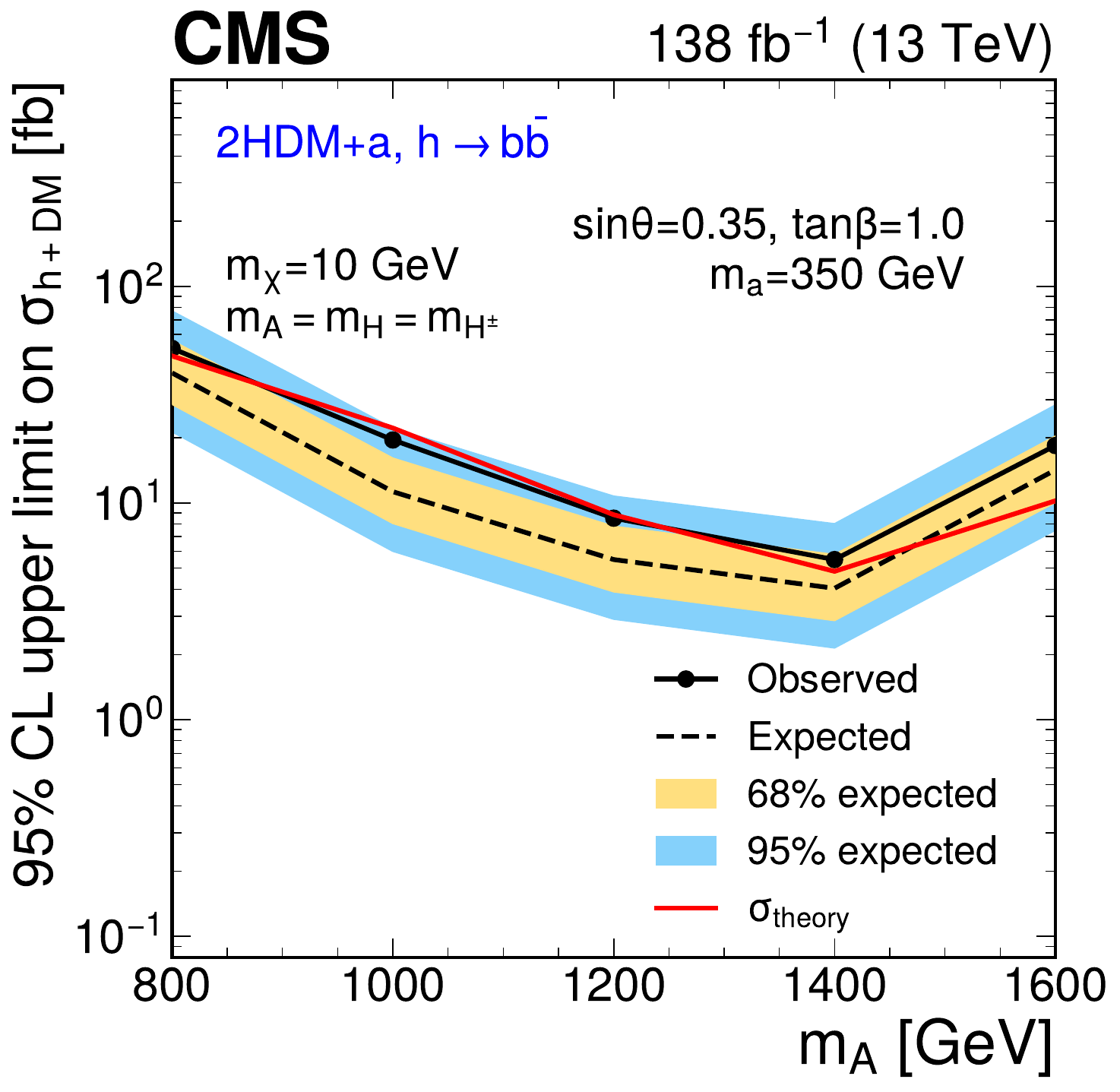} \\
  \includegraphics[width=0.45\textwidth]{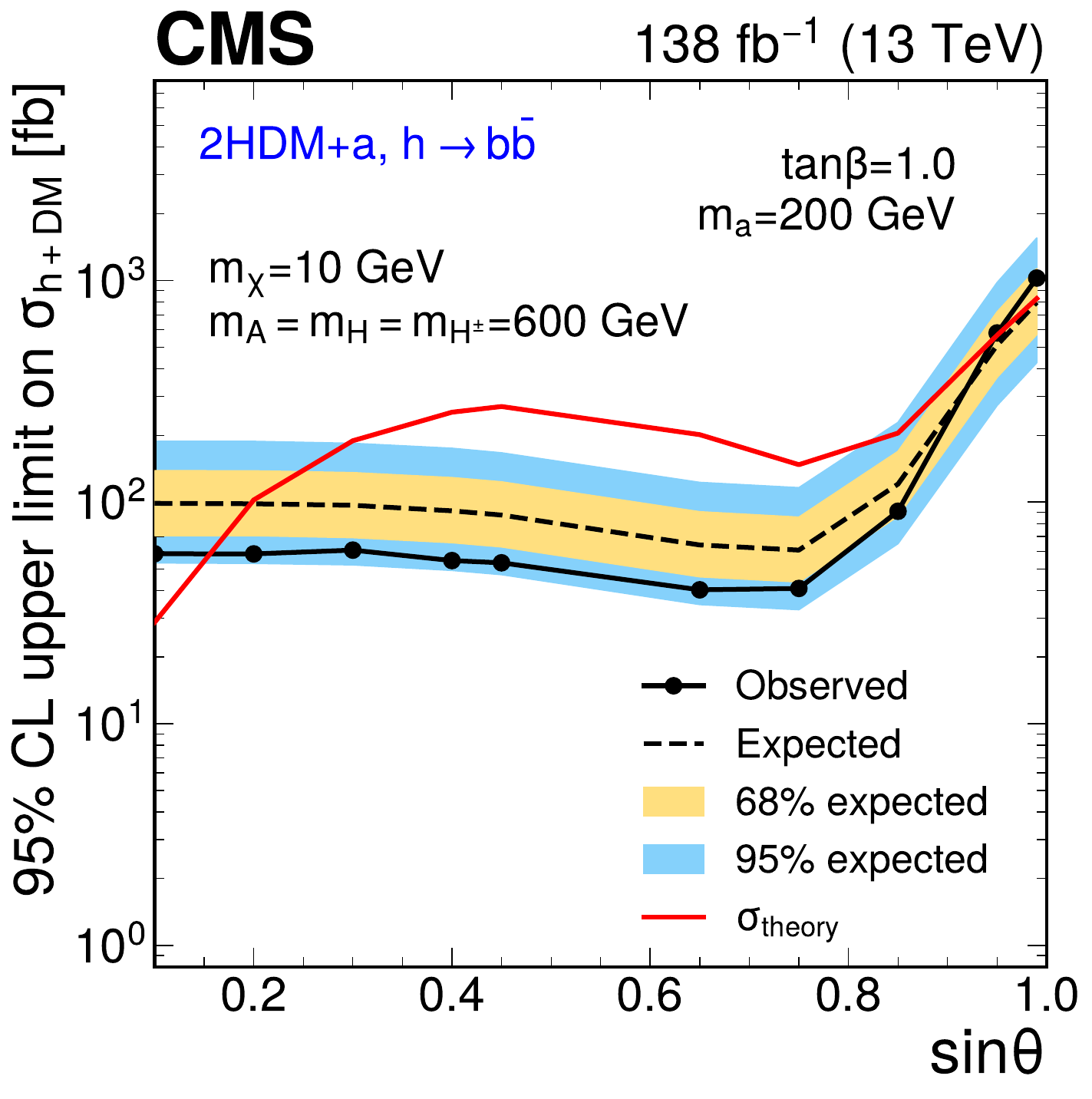}
  \hfill%
  \includegraphics[width=0.45\textwidth]{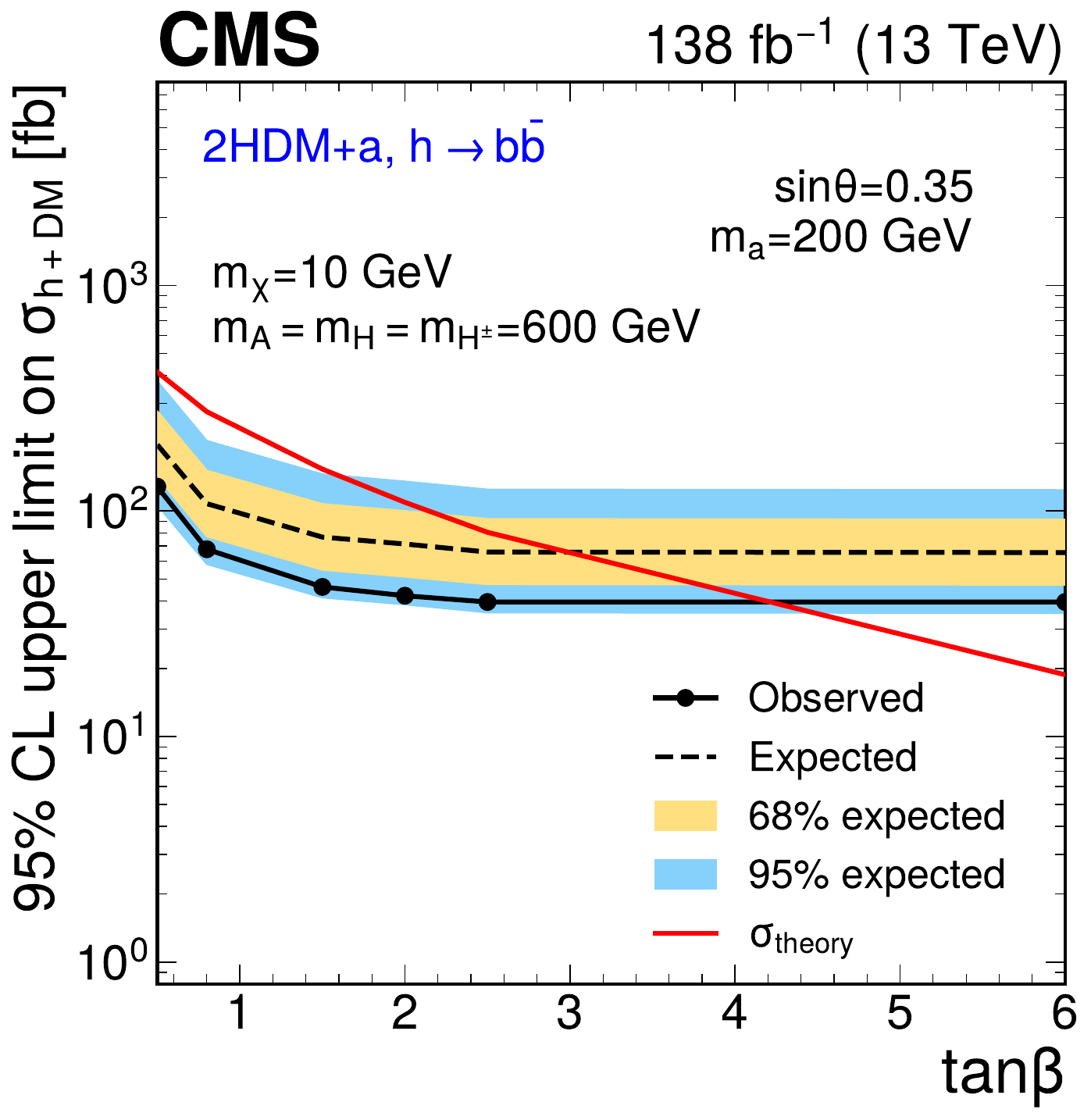}
  \caption{Observed and expected exclusion limits at 95\% \CL on the signal cross section \sighDM for the \twohdma model as a function of the model parameters: \ma (upper left), \mA (upper right), \sintheta (lower left), and \tanbeta (lower right) while fixing the values of the other parameters, as indicated in the legends. Different sets of model parameters are tested to probe distinct regions of phase space. Mass points below the solid red line are excluded.}
  \label{limits:2HDMa}
\end{figure*}

Figure~\ref{limits:2HDMa} shows the observed and expected exclusion limits at 95\% \CL on the signal cross section \sighDM for four one-dimensional parameter scans (\ma, \mA, \sintheta, and \tanbeta). For the \ma scan (upper left plot), the fixed parameters are $\mA=1000\GeV$, $\sintheta=0.35$, and $\tanbeta=1$. Light pseudoscalar masses below 360\GeV are excluded. For the \mA scan (upper right plot), the fixed parameters are $\ma=350\GeV$, $\sintheta=0.35$, and $\tanbeta=1$, with heavy pseudoscalar masses between 850 and 1300\GeV excluded. For the \sintheta scan (lower left plot), the fixed parameters are $\ma=200\GeV$, $\mA=600\GeV$, and $\tanbeta=1$, with \sintheta values between 0.15 and 0.95 excluded. The variation of \sintheta leads to both rate and shape changes in the \ptmiss spectrum, where the strong sensitivity is a result of the interplay of resonant and nonresonant contributions. The process $\pp\to\PSA\to\PSh\Pa\to\PSh\chichibar$ scales as $\sintheta\cossquaretheta$ and the process $\pp\to\PSh\Pa\to\PSh\chichibar$ shows a \sintheta dependence, as detailed in Ref.~\cite{abe2018lhc}. Finally, for the \tanbeta scan (lower right plot), the fixed parameters are $\ma=200\GeV$, $\mA=600\GeV$, and $\sintheta=0.35$, where \tanbeta values below 4.2 are excluded by this analysis. As long as $\mchi<\ma/2$, the decay is kinematically allowed, with very little dependence of either the rate or \ptmiss spectrum on \mchi. We choose a benchmark value $\mchi=10\GeV$~\cite{abe2018lhc}.

{\tolerance=800
Motivated by Ref.~\cite{abe2018lhc}, different sets of model parameters are tested, corresponding to $(\mA,\ma)=(1000,350)\GeV$ (Fig.~\ref{limits:2HDMa} upper plots) and $(600,200)\GeV$ (Fig.~\ref{limits:2HDMa} lower plots). These choices probe distinct regions of phase space, resulting in signal events with different kinematic features. In particular, the former configuration yields higher \ptmiss, and together with different production cross sections, these variations result in differences between the observed and expected exclusion limits.
\par}

\section{Summary}
\label{sec:Summary}

A search for DM produced in association with a standard model Higgs boson decaying to a bottom quark-antiquark pair has been presented. The analysis is based on proton-proton collision data collected in 2017 and 2018 by the CMS experiment at $\sqrt{s}=13\TeV$, corresponding to an integrated luminosity of 101\fbinv. The search has been performed in merged and resolved categories to cover a wide range of Lorentz boosts of the Higgs boson. The signal is extracted using a simultaneous fit to the signal and control regions by combining the two categories. The observed data agree with the standard model background prediction, indicating no evidence for new physics. The results are statistically combined with an earlier search using 2016 data, corresponding to an integrated luminosity of 35.9\fbinv.

The full 2016--2018 results have been interpreted in two simplified models, the \baryonicZprime model where a high mass resonance (\PZpr) decays to a pair of DM particles and a Higgs boson, and the \twohdma model where a heavy pseudoscalar couples to a Higgs boson and a lighter pseudoscalar that decays to a pair of DM particles. Exclusion limits are set on the model parameters at 95\% confidence level. For the \baryonicZprime model, mediator \PZpr masses up to 2.25\TeV are excluded for a DM mass of 1\GeV, and DM particle masses up to 550\GeV are excluded for a 1.25\TeV \PZpr particle. In the \twohdma framework, light pseudoscalar masses \ma below 360\GeV are excluded for a heavy pseudoscalar mass \mA of 1000\GeV, and \mA masses between 850 and 1300\GeV are excluded for \ma of 350\GeV. For the other model parameters, \sintheta values between 0.15 and 0.95 are excluded, while \tanbeta values less than 4.2 are excluded. These results improve upon the previously existing CMS limits owing to the larger integrated luminosity and improved identification of $\PSh\to\bbbar$ decay.

\begin{acknowledgments}
We congratulate our colleagues in the CERN accelerator departments for the excellent performance of the LHC and thank the technical and administrative staffs at CERN and at other CMS institutes for their contributions to the success of the CMS effort. In addition, we gratefully acknowledge the computing centers and personnel of the Worldwide LHC Computing Grid and other centers for delivering so effectively the computing infrastructure essential to our analyses. Finally, we acknowledge the enduring support for the construction and operation of the LHC, the CMS detector, and the supporting computing infrastructure provided by the following funding agencies: SC (Armenia), BMBWF and FWF (Austria); FNRS and FWO (Belgium); CNPq, CAPES, FAPERJ, FAPERGS, and FAPESP (Brazil); MES and BNSF (Bulgaria); CERN; CAS, MoST, and NSFC (China); MINCIENCIAS (Colombia); MSES and CSF (Croatia); RIF (Cyprus); SENESCYT (Ecuador); ERC PRG, TARISTU24-TK10 and MoER TK202 (Estonia); Academy of Finland, MEC, and HIP (Finland); CEA and CNRS/IN2P3 (France); SRNSF (Georgia); BMFTR, DFG, and HGF (Germany); GSRI (Greece); NKFIH (Hungary); DAE and DST (India); IPM (Iran); SFI (Ireland); INFN (Italy); MSIT and NRF (Republic of Korea); MES (Latvia); LMTLT (Lithuania); MOE and UM (Malaysia); BUAP, CINVESTAV, CONACYT, LNS, SEP, and UASLP-FAI (Mexico); MOS (Montenegro); MBIE (New Zealand); PAEC (Pakistan); MES, NSC, and NAWA (Poland); FCT (Portugal);  MESTD (Serbia); MICIU/AEI and PCTI (Spain); MOSTR (Sri Lanka); Swiss Funding Agencies (Switzerland); MST (Taipei); MHESI (Thailand); TUBITAK and TENMAK (T\"{u}rkiye); NASU (Ukraine); STFC (United Kingdom); DOE and NSF (USA).

\hyphenation{Rachada-pisek} Individuals have received support from the Marie-Curie program and the European Research Council and Horizon 2020 Grant, contract Nos.\ 675440, 724704, 752730, 758316, 765710, 824093, 101115353, 101002207, 101001205, and COST Action CA16108 (European Union); the Leventis Foundation; the Alfred P.\ Sloan Foundation; the Alexander von Humboldt Foundation; the Science Committee, project no. 22rl-037 (Armenia); the Fonds pour la Formation \`a la Recherche dans l'Industrie et dans l'Agriculture (FRIA) and Fonds voor Wetenschappelijk Onderzoek contract No. 1228724N (Belgium); the Beijing Municipal Science \& Technology Commission, No. Z191100007219010, the Fundamental Research Funds for the Central Universities, the Ministry of Science and Technology of China under Grant No. 2023YFA1605804, the Natural Science Foundation of China under Grant No. 12535004, and USTC Research Funds of the Double First-Class Initiative No.\ YD2030002017 (China); the Ministry of Education, Youth and Sports (MEYS) of the Czech Republic; the Shota Rustaveli National Science Foundation, grant FR-22-985 (Georgia); the Deutsche Forschungsgemeinschaft (DFG), among others, under Germany's Excellence Strategy -- EXC 2121 ``Quantum Universe" -- 390833306, and under project number 400140256 - GRK2497; the Hellenic Foundation for Research and Innovation (HFRI), Project Number 2288 (Greece); the Hungarian Academy of Sciences, the New National Excellence Program - \'UNKP, the NKFIH research grants K 131991, K 133046, K 138136, K 143460, K 143477, K 146913, K 146914, K 147048, 2020-2.2.1-ED-2021-00181, TKP2021-NKTA-64, and 2025-1.1.5-NEMZ\_KI-2025-00004 (Hungary); the Council of Science and Industrial Research, India; ICSC -- National Research Center for High Performance Computing, Big Data and Quantum Computing, FAIR -- Future Artificial Intelligence Research, and CUP I53D23001070006 (Mission 4 Component 1), funded by the NextGenerationEU program (Italy); the Latvian Council of Science; the Ministry of Education and Science, project no. 2022/WK/14, and the National Science Center, contracts Opus 2021/41/B/ST2/01369, 2021/43/B/ST2/01552, 2023/49/B/ST2/03273, and the NAWA contract BPN/PPO/2021/1/00011 (Poland); the Funda\c{c}\~ao para a Ci\^encia e a Tecnologia, grant CEECIND/01334/2018 (Portugal); the National Priorities Research Program by Qatar National Research Fund; MICIU/AEI/10.13039/501100011033, ERDF/EU, ``European Union NextGenerationEU/PRTR", and Programa Severo Ochoa del Principado de Asturias (Spain); the Chulalongkorn Academic into Its 2nd Century Project Advancement Project, the National Science, Research and Innovation Fund program IND\_FF\_68\_369\_2300\_097, and the Program Management Unit for Human Resources \& Institutional Development, Research and Innovation, grant B39G680009 (Thailand); the Kavli Foundation; the Nvidia Corporation; the SuperMicro Corporation; the Welch Foundation, contract C-1845; and the Weston Havens Foundation (USA).  
\end{acknowledgments}

\bibliography{auto_generated}
\cleardoublepage \appendix\section{The CMS Collaboration \label{app:collab}}\begin{sloppypar}\hyphenpenalty=5000\widowpenalty=500\clubpenalty=5000
\cmsinstitute{Yerevan Physics Institute, Yerevan, Armenia}
{\tolerance=6000
A.~Hayrapetyan, V.~Makarenko\cmsorcid{0000-0002-8406-8605}, A.~Tumasyan\cmsAuthorMark{1}\cmsorcid{0009-0000-0684-6742}
\par}
\cmsinstitute{Institut f\"{u}r Hochenergiephysik, Vienna, Austria}
{\tolerance=6000
W.~Adam\cmsorcid{0000-0001-9099-4341}, L.~Benato\cmsorcid{0000-0001-5135-7489}, T.~Bergauer\cmsorcid{0000-0002-5786-0293}, M.~Dragicevic\cmsorcid{0000-0003-1967-6783}, C.~Giordano, P.S.~Hussain\cmsorcid{0000-0002-4825-5278}, M.~Jeitler\cmsAuthorMark{2}\cmsorcid{0000-0002-5141-9560}, N.~Krammer\cmsorcid{0000-0002-0548-0985}, A.~Li\cmsorcid{0000-0002-4547-116X}, D.~Liko\cmsorcid{0000-0002-3380-473X}, M.~Matthewman, J.~Schieck\cmsAuthorMark{2}\cmsorcid{0000-0002-1058-8093}, R.~Sch\"{o}fbeck\cmsAuthorMark{2}\cmsorcid{0000-0002-2332-8784}, D.~Schwarz\cmsorcid{0000-0002-3821-7331}, M.~Shooshtari\cmsorcid{0009-0004-8882-4887}, M.~Sonawane\cmsorcid{0000-0003-0510-7010}, W.~Waltenberger\cmsorcid{0000-0002-6215-7228}, C.-E.~Wulz\cmsAuthorMark{2}\cmsorcid{0000-0001-9226-5812}
% ...
\par}
\vskip\cmsinstskip
\dag:~Deceased\\
$^{1}$Also at Yerevan State University, Yerevan, Armenia\\
$^{2}$Also at TU Wien, Vienna, Austria\\
% ...
\end{sloppypar}
\end{document}